\documentclass[acmsmall,screen]{acmart}

\AtBeginDocument{%
  }

\usepackage{caption}
\usepackage{subcaption}

\newcommand{\editmade}[1]{#1}

\setcopyright{usgovmixed}
\copyrightyear{2022}
\acmYear{2022}
\acmDOI{XXXXXXX.XXXXXXX}

\acmJournal{TOMS}
\acmVolume{0}
\acmNumber{0}
\acmArticle{0}
\acmMonth{0}

\newcommand{\R}{\mathbb{R}}

\newcommand{\bigO}[1]{\ensuremath{\mathop{}\mathopen{}\mathcal{O}\mathopen{}\left(#1\right)}}

\begin{document}

\title[The ARKODE IVP solver infrastructure]{ARKODE: A flexible IVP solver infrastructure for one-step methods}


\author{DANIEL R. REYNOLDS}
\orcid{0000-0002-0911-7841}
\affiliation{%
    \institution{Southern Methodist University}
    \department{Department of Mathematics}
    \city{Dallas}
    \state{Texas}
    \country{USA}
    \postcode{75275}}
\email{reynolds@smu.edu}

\author{DAVID J. GARDNER}
\orcid{0000-0002-7993-8282}
\affiliation{%
    \institution{Lawrence Livermore National Laboratory}
    \department{Center for Applied Scientific Computing}
    \city{Livermore}
    \state{California}
    \country{USA}
}
\email{gardner48@llnl.gov}

\author{CAROL S. WOODWARD}
\orcid{0000-0002-6502-8659}
\affiliation{%
    \institution{Lawrence Livermore National Laboratory}
    \department{Center for Applied Scientific Computing}
    \city{Livermore}
    \state{California}
    \country{USA}
  }
\email{woodward6@llnl.gov}

\author{RUJEKO CHINOMONA}
\orcid{0000-0002-5267-8433}
\affiliation{%
    \institution{Temple University}
    \department{Department of Mathematics}
    \city{Philadelphia}
    \state{Pennsylvania}
    \country{USA}
  }
\email{rchinomona@temple.edu}
\renewcommand{\shortauthors}{D.~R.~Reynolds, D.~J.~Gardner, C.~S.~Woodward, and R.~Chinomona}

\authorsaddresses{%
Author's addresses: D.~R.~Reynolds, reynolds@smu.edu, Department of Mathematics, Southern Methodist University; D.~J.~Gardner, gardner48@llnl.gov; C.~S.~Woodward, woodward6@llnl.gov, Center for Applied Scientific Computing, Lawrence Livermore National Laboratory; and R.~Chinomona, rujeko.chinomona@temple.edu, Department of Mathematics, Temple University.}

\begin{abstract}
We describe the ARKODE library of one-step time integration methods for \editmade{ordinary differential equation (ODE)} initial-value problems (IVPs). In addition to providing standard explicit and diagonally implicit Runge--Kutta methods, ARKODE also supports one-step methods designed to treat additive splittings of the IVP, including implicit-explicit (ImEx) additive Runge--Kutta methods and multirate infinitesimal (MRI) methods.  We present the role of ARKODE within the SUNDIALS suite of time integration and nonlinear solver libraries, the core ARKODE infrastructure for utilities common to large classes of one-step methods, as well as its use of ``time stepper'' modules enabling easy incorporation of novel algorithms into the library.  Numerical results show example problems of increasing complexity, highlighting the algorithmic flexibility afforded through this infrastructure, and \editmade{include} a larger multiphysics application leveraging multiple algorithmic features from ARKODE and SUNDIALS.
\end{abstract}



\maketitle

\section{Introduction}
\label{sec:intro}

The SUite of Nonlinear and DIfferential/ALgebraic Solvers (SUNDIALS) \cite{HBGLSSW2005,gardner2020enabling,balosEnablingGPUAccelerated2021} is a numerical software library providing time integrators and nonlinear solvers for use on a range of computing systems -- from
laptops to leadership class supercomputers.
The newest package in SUNDIALS, ARKODE, provides highly flexible time integration methods for additive systems with partitions of differing stiffness (implicit-explicit methods) and differing time scales (multirate methods).  ARKODE focuses on one-step, multi-stage methods for first-order initial-value problems (IVPs) that typically include robust temporal adaptivity, based on estimates of both solution error and solver efficiency.  These methods leverage significant theoretical developments over recent decades in Runge--Kutta methods, and they naturally map to space-time systems of partial differential equations (PDEs) that employ spatially-adaptive meshes.  Furthermore, ARKODE can be considered as an infrastructure for one-step methods, providing a range of reusable components that may be replaced with problem-optimal choices, or even experimental methods for rapid development and testing.  The ARKODE implementations for different classes of methods follow similar APIs and leverage many heuristics from other SUNDIALS packages allowing for similar usage and efficiency.

Over the last two decades, there have been significant advances in the derivation of numerical methods that allow both high accuracy and increased flexibility with regard to how various components of a problem are treated.  These methods range from those that apply a uniform time step size for an entire problem while varying the algorithms used on individual components, to multirate methods that evolve separate solution components using different step sizes.

Methods in the former category have been introduced primarily for problems that couple stiff and nonstiff processes. Instead of employing a fully implicit or fully explicit method that would be ideally suited to only the stiff or nonstiff components of the problem, respectively, these approaches apply robust implicit solvers to the stiff components while treating the remaining nonstiff (and frequently nonlinear) components explicitly.  While simplistic first-order operator-split approaches have been utilized for decades, including higher-order variants that introduce complex arithmetic or backward integration \cite{cervi2019comparison}, novel methods that provide increased accuracy and stability for such problems include implicit-explicit (ImEx) additive Runge--Kutta methods, first introduced in \cite{cooperAdditiveMethodsNumerical1980} with newer embedded versions supporting temporal adaptivity developed in \cite{kennedyAdditiveRungeKutta2003,kennedyHigherorderAdditiveRunge2019}.

Multirate methods, on the other hand, evolve separate problem components using different time step sizes and are frequently applied to multiphysics problems that combine various physical processes which may separately evolve on disparate time scales.  In such circumstances the ``fast'' processes are often evolved with small step sizes, while the ``slow'' processes are evolved using larger time steps.  Here again, simple first-order ``subcycling'' approaches have been employed for many years; however, research into higher-order approaches has recently seen dramatic advances through development of multirate methods \cite{constantinescuExtrapolatedMultirateMethods2013,gearMultirateLinearMultistep1984}
including multirate infinitesimal methods \cite{KW1998, schlegelMultirateRungeKutta2009,wenschMultirateInfinitesimalStep2009,sanduClassMultirateInfinitesimal2019,robertsCoupledMultirateInfinitesimal2020,luanNewClassHighOrder2020,CR2021} and
multirate generalized additive Runge-Kutta methods \cite{guntherMultirateGeneralizedAdditive2016,roberts2021implicit}.

Recently, several software packages have been developed to meet some of the challenges presented by \editmade{time-dependent} multiphysics problems.  Some of the most notable ones include
the PETSc/TS \cite{ABCGSZ2018} and Trilinos/Tempus \cite{tempus-website} packages, that respectively provide C and C++ implementations of explicit, implicit, and additive Runge--Kutta methods for
ODEs and DAEs.  Additionally, the DifferentialEquations.jl package \cite{rackauckas2017differentialequations} provides a suite for numerically solving ordinary differential equations and stochastic differential equations (including back-end interfaces to SUNDIALS) written in Julia, Python, and R.  All of these packages include temporal adaptivity and additive formulations that support splitting a problem into its stiff and nonstiff components.  Additionally, these packages provide support for solving the nonlinear and linear systems of equations that arise from implicit time discretizations, as well as implementations that enable computations on GPU accelerators.  \editmade{Of these packages, multirate methods}, which are critical for a wide range of multiphysics applications, \editmade{are only supported by PETSc, which includes second-order Multirate Partitioned Runge--Kutta methods \cite{KangEtAl2022}}.

The goal of the ARKODE solver library is to provide a bridge between recent mathematical advances into flexible time integration methods and the large-scale simulations that need them.  Our overall aims in designing ARKODE were three-fold:
\begin{enumerate}
\item Provide high-order, stable, efficient, and scalable methods for multiphysics, multirate systems of ordinary differential equation (ODE) IVPs, for which no single ``textbook'' solution technique optimally applies to all of its various components.
\item Provide both an IVP solver and experimentation framework that allows users to rapidly explore algorithmic optimizations.
\item Provide a high-quality, scalable, open-source software library that can easily be incorporated into existing simulation codes \emph{without} requiring application developers to rewrite or convert their native data structures, and that can be run on a broad range of computing systems.
\end{enumerate}

While ARKODE may be used as a standalone library, it is released as part of, and leverages numerous components from, the larger SUNDIALS suite, which also consists of the solvers CVODE (for ODE IVP systems), IDA (for differential-algebraic IVP systems), CVODES and IDAS (forward and adjoint sensitivity analysis variants of CVODE and IDA, respectively), and KINSOL (for nonlinear algebraic systems). As part of SUNDIALS, ARKODE has access to a rich ecosystem of vector, linear solver, and nonlinear solver classes to foster both experimentation \emph{between} SUNDIALS integrators (e.g., CVODE vs ARKODE) and to leverage existing interfaces and infrastructure.  SUNDIALS, including ARKODE, is written in C with modern Fortran interfaces.  All operations on data within the SUNDIALS packages are done through a clear set of abstract vector, matrix, linear solver, and nonlinear solver interfaces.  To wit, SUNDIALS integrators, including ARKODE, make no assumption about how a user's data is laid out in memory; specifically, users can (if they wish) supply their own class implementations, as long as they provide the methods needed by the integrator.

This flexibility allows for ARKODE, along with other SUNDIALS integrators, to be easily employed from within existing applications. For example, ARKODE has been used in several applications including the High-Order Methods Modeling Environment (HOMME) dynamical core in the Energy Exascale Earth System Model (E3SM) \cite{vogl2019evaluation}, the Tempest atmospheric dynamical core \cite{gardner2018implicit}, the ParaDiS dislocation dynamics simulator \cite{gardner2015implicit}, the Modular Finite Element Methods Library (MFEM) \cite{mfem}, the PeleC adaptive-mesh compressible hydrodynamics code for reacting flows \cite{sitaraman2021adaptive}, the Mushroom fusion code \cite{mushroom-aps-2021}, and the MEUMAPPS phase-field modeling code \cite{radhakrishnan2016phase, doecode_69332}.

ARKODE also serves as an infrastructure for one-step methods, and, as such, it provides a variety of shared modules for (a) handling temporal adaptivity to achieve a desired solution accuracy and efficiently utilize any underlying iterative algebraic (nonlinear and linear) solvers, (b) interfaces to translate user-defined IVP right-hand side and Jacobian routines into the routines required by SUNDIALS' general purpose algebraic solver classes, (c) stiff and nonstiff temporal interpolation modules for dense output and efficient implicit predictors for iterative algebraic solvers, (d) root-finding capabilities for event detection, and (e) enforcing solution inequality constraints.

The rest of this paper is organized as follows.  In Section \ref{sec:timesteppers} we introduce the current time stepping modules provided within ARKODE, as these user-facing components define both the types of problems that may be solved and the algorithms that can be applied to them. For readers interested in the underlying structure of ARKODE, in Section \ref{sec:infrastructure} we discuss the shared infrastructure on which ARKODE's time stepping modules are built.  Then, in Section \ref{sec:usage}, we present the standard usage of ARKODE. In Section \ref{sec:demonstration} we present a sequence of simple numerical examples highlighting the algorithmic flexibility of ARKODE and a demonstration example showing an advanced use of ARKODE for a large-scale multiphysics application. Finally, in Section \ref{sec:conclusions} we point out open-source options for accessing ARKODE and provide concluding remarks.  For the sake of readability, we omit some specific usage and implementation details; for this information, as well as additional details on ARKODE itself, we refer readers to the ARKODE documentation at \href{https://sundials.readthedocs.io/}{sundials.readthedocs.io}.

\section{Time-stepping Modules}
\label{sec:timesteppers}

ARKODE considers time-dependent initial-value problems in linearly implicit first order form,
\begin{align}
  \label{eq:IVP}
  M(t)\, y'(t) = f(t,y),\qquad y(t_0) = y_0,
\end{align}
where the independent variable typically satisfies $t\in[t_0,t_f]$ (ARKODE also allows \editmade{$y(t_f)=y_f$}, in which case \eqref{eq:IVP} is a \editmade{terminal}-value problem), and the dependent variable is $y\in\R^N$. $M(t)$ is a user-specified nonsingular operator from $\R^N\to\R^N$, that is assumed to be independent of $y$.  For standard systems of ODEs and for problems arising from the spatial semi-discretization of PDEs using finite difference, finite volume, or spectral element methods, $M$ is typically the identity matrix, $I$. For PDEs using finite-element spatial semi-discretizations, $M$ is \editmade{(or can be taken to be)} a well-conditioned mass matrix.

While each of the time-stepping modules supplied with ARKODE considers different formulations of the problem \eqref{eq:IVP}, all share the basic structure wherein $y$ is provided at only the initial (or final) time, and ARKODE computes approximate values of $y(t)$ at a discrete set of independent variable values, either $t_0 < t_1 < \ldots < t_f$ for forward-in-time evolution, or $t_f > \ldots > t_1 > t_0$ for reverse evolution.  We denote these approximations of the solution $y(t_n)$ as $y_n$.  One-step methods for solution of \eqref{eq:IVP} generate these approximate solutions using some formula,
\begin{align}
  \label{eq:one_step_method}
  y_n = \varphi(t_{n-1}, y_{n-1}, y_n, h_n),
\end{align}
where $h_n = t_n-t_{n-1}$ is the step size, and $\varphi$ denotes the approximation method that updates the solution $y_{n-1} \to y_n$.  ARKODE makes no assumption that the step sizes are uniform, i.e. $h_n \ne h_{n-1}$, although such use is supported.  Instead, ARKODE strives to take the largest possible step that simultaneously ensures that the approximations, $y_n$, are sufficiently accurate and that the time-stepping algorithms used to evaluate $\varphi$ are efficient and robust.

\begin{figure}[htb!]
    \centering
    \includegraphics[width=0.8\textwidth]{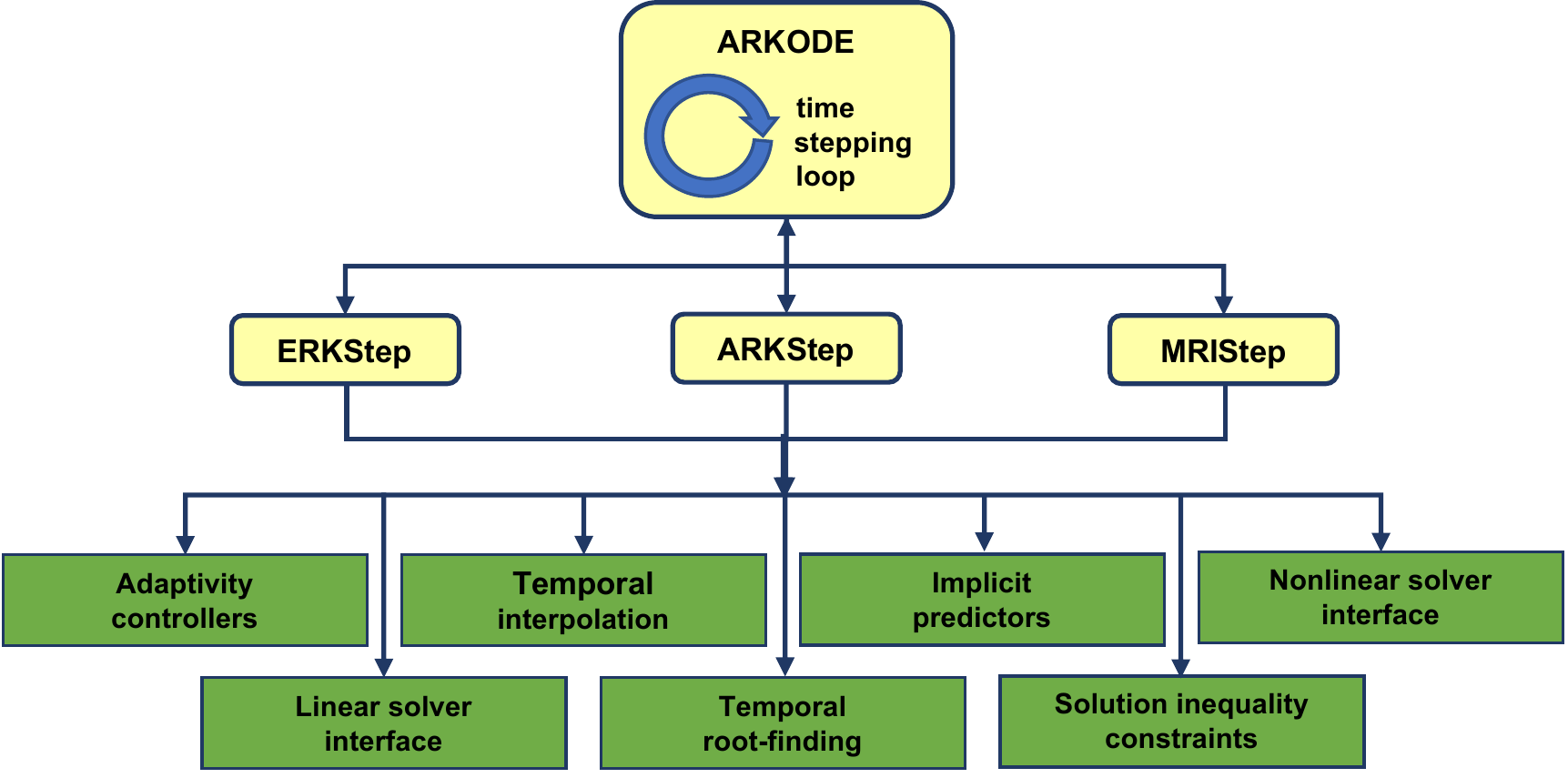}
    \caption{The relationship between ARKODE and its time-stepping modules.  ARKODE provides the time integration loop; each stepper performs individual time steps leveraging the shared infrastructure as needed.}
    \label{fig:stepper-diagram}
\end{figure}

While the ARKODE infrastructure provides utilities for temporal integration of one-step IVP methods, it does not specify either the \emph{type} of IVP problems that should be solved (i.e., the structure of the function $f(t,y)$), nor does it require any specific numerical algorithm for taking the step, $\varphi(t_{n-1}, y_{n-1}, y_n, h_n)$.  These aspects of integration are left up to specific \emph{time stepping modules} that rest on the ARKODE infrastructure -- a diagram of this relationship is shown in Figure \ref{fig:stepper-diagram}.  The time-stepping modules provide the user-facing software layer for problem definition, evolution, and reporting of solver statistics.  Upon creation of a time-stepping module, that module internally creates the main ARKODE memory structure and ``attaches'' itself to ARKODE by providing two functions used by ARKODE during the IVP evolution:
\begin{enumerate}
\item an \emph{initialization} function that is called by ARKODE once all setup has completed and temporal integration of the problem is about to commence, and
\item a \emph{step} function that computes a candidate update, $\tilde{y} = \varphi(t_{n-1}, y_{n-1}, h_n)$, and the corresponding estimate of the local truncation error, $\|T_n\|$.
\end{enumerate}

Currently, ARKODE includes three time-stepping modules: ERKStep, ARKStep, and MRIStep, discussed in the following subsections.

\subsection{ERKStep}
\label{sec:erkstep}

The ERKStep time-stepping module is designed for IVPs of the form
\begin{equation}
  \label{eq:ERKStep_IVP}
  y'(t) = f(t,y), \qquad y(t_0) = y_0,
\end{equation}
i.e., $M(t)=I$ in \eqref{eq:IVP}.  For such problems, ERKStep provides variable-step, embedded, \emph{explicit Runge-Kutta} (ERK) methods, that evolve a single time step $t_{n-1}\to t_{n-1}+h_n = t_n$ using the algorithm
\begin{subequations}
  \label{eq:ERK_method}
  \begin{align}
    \label{eq:ERK_stage}
    z_i &= y_{n-1} + h_n \sum_{j=1}^{i-1} A_{i,j} f(t_{n,j}, z_j),
      \quad i=1,\ldots,s, \\
    y_n &= y_{n-1} + h_n \sum_{i=1}^{s} b_i f(t_{n,i}, z_i), \\
    \tilde{y}_n &= y_{n-1} + h_n \sum_{i=1}^{s} \tilde{b}_i f(t_{n,i}, z_i),
  \end{align}
\end{subequations}
where the internal stage times are abbreviated using the notation, $t_{n,j}=t_{n-1}+c_jh_n$, and the coefficients $A\in\R^{s\times s}$, $b\in\R^s$, $c\in\R^s$, and optional embedding coefficients $\tilde{b}\in\R^s$ (a.k.a., the Butcher table) define the method. The embedding,
$\tilde{y}_n$. typically provides a slightly lower accuracy approximation than the computed solution, $y_n$.
ARKODE provides a number of ERK Butcher tables having orders of accuracy $q=\{2,3,4,5,6,8\}$, each is supplied with an embedding $\tilde{b}$ that usually has order of accuracy $p=q-1$.
User-supplied Butcher tables are also supported and are not required to include embedding coefficients, although if these are not provided then automatic temporal adaptivity will be disabled.  We have found this ``fixed-step'' mode to be particularly useful when existing simulation codes transition to ARKODE.  By providing the ERK Butcher table that a code currently uses and ceding control over $h_n$ to the application code, users may run verification tests to ensure reproducibility before switching to higher-order and/or adaptive ERK methods.

In order to allow users to ``tune'' ERKStep to better exploit problem structure, numerous optional features may be enabled or modified by the user.  The full set of features is described in \cite{arkodeUserGuide}; a few of which are highlighted in the example problems \editmade{in Section \ref{sec:erkstep_example}}.

\subsection{ARKStep}
\label{sec:arkstep}

The ARKStep time-stepping module is designed for IVPs where the right-hand side function may be additively split into two components,
\begin{equation}
  \label{eq:ARKStep_IVP}
  M(t)\, y'(t) = f^E(t,y) + f^I(t,y), \qquad y(t_0) = y_0,
\end{equation}
where the user-provided right-hand side function $f^E(t,y)$ contains the ``nonstiff'' components of the system, and $f^I(t,y)$ contains the ``stiff'' components of the system.
In solving \eqref{eq:ARKStep_IVP}, we first convert to standard additive form,
\begin{equation}
  \label{eq:ARKStep_IVP_standard}
  y'(t) = \hat{f}^E(t,y) + \hat{f}^I(t,y), \qquad y(t_0) = y_0,
\end{equation}
where $\hat{f}^E(t,y) = M(t)^{-1}f^E(t,y)$ and $\hat{f}^I(t,y) = M(t)^{-1}f^I(t,y)$.  ARKStep utilizes variable-step, embedded, ImEx \emph{additive Runge-Kutta} (ARK) methods.  Given a pair of $s$-stage Butcher tables $(A^E, b^E, \tilde{b}^E, c^E)$ and $(A^I, b^I, \tilde{b}^I, c^I)$, a single time step $t_{n-1}\to t_{n}$ is advanced via the algorithm
\begin{subequations}
  \label{eq:ARK_method}
  \begin{align}
    \label{eq:ARK_stage}
    z_i &= y_{n-1} + h_n \sum_{j=1}^{i-1} A^E_{i,j} \hat{f}^E(t^E_{n,j}, z_j)
      + h_n \sum_{j=1}^{i} A^I_{i,j} \hat{f}^I(t^I_{n,j}, z_j), \quad i=1,\ldots,s, \\
    y_n &= y_{n-1} + h_n \sum_{i=1}^{s} \left(b^E_i \hat{f}^E(t^E_{n,i}, z_i)
      + b^I_i \hat{f}^I(t^I_{n,i}, z_i)\right), \\
    \tilde{y}_n &= y_{n-1} + h_n \sum_{i=1}^{s} \left(
      \tilde{b}^E_i \hat{f}^E(t^E_{n,i}, z_i) +
      \tilde{b}^I_i \hat{f}^I(t^I_{n,i}, z_i)\right),
  \end{align}
\end{subequations}
where we denote the internal stage times as $t^E_{n,j} = t_{n-1} + c^E_j h_n$ and $t^I_{n,j} = t_{n-1} + c^I_j h_n$.
We note that ARKStep enforces the constraint that the explicit and implicit Butcher tables share the same number of stages, $s$, but it allows for the possibility of different explicit and implicit abscissae, i.e., $c^E \neq c^I$ (as required for various SSP ImEx-ARK methods \cite{H2006,H2009,HHKK2014,PR2005}).

Through appropriate selection of the Butcher tables and right-hand side functions, ARKStep supports three classes of Runge--Kutta methods: \emph{ImEx}, \emph{explicit}, and \emph{diagonally implicit}. Several Butcher tables are provided for each class, all of which include the embedding coefficients, $\tilde{b}^E$ and $\tilde{b}^I$, although users may provide their own tables without embeddings (in which case adaptivity will be disabled). For mixed stiff/nonstiff problems, a user should provide both of the functions $f^E$ and $f^I$ that define the IVP system, in which case ARKStep provides the ARK Butcher tables proposed in \cite{KC2003,KC2019} with orders of accuracy $q = \{3,4,5\}$ and embeddings of orders $p = \{2,3,4\}$, respectively.

Either of the function pointers for $f^I$ or $f^E$ may be \texttt{NULL} to disable that component.  For nonstiff problems one can disable $f^I$, in which case \eqref{eq:ARKStep_IVP} reduces to the non-split IVP
\begin{equation}
  \label{eq:ARKStep_IVP_explicit}
  M(t)\, y'(t) = f^E(t,y), \qquad y(t_0) = y_0.
\end{equation}
In this scenario, the implicit Butcher table ($A^I$, $b^I$, $\tilde{b}^I$, $c^I$) in \eqref{eq:ARK_method} is ignored, and the ARK methods reduce to classical ERK methods.  Here, all methods from ERKStep are again available in ARKStep.
We note that in this mode, both the problem \eqref{eq:ARKStep_IVP_explicit} and the method \eqref{eq:ARK_method} fully encapsulate the ERKStep problem \eqref{eq:ERKStep_IVP} and method \eqref{eq:ERK_method}.  While it therefore follows that ARKStep can be used to solve every problem solvable by ERKStep (even with the same Butcher tables), we retain ERKStep as a distinct time-stepping module since its simplified form admits a more efficient and memory-friendly implementation than that provided by ARKStep.

Alternately, for stiff problems $f^E$ may be disabled so that \eqref{eq:ARKStep_IVP} reduces to
\begin{equation}
  \label{eq:ARKStep_IVP_implicit}
  M(t)\, y'(t) = f^I(t,y), \qquad y(t_0) = y_0.
\end{equation}
Here the explicit Butcher table ($A^E$, $b^E$, $\tilde{b}^E$, $c^E$) in \eqref{eq:ARK_method} is ignored, and the ARK methods reduce to standard \emph{diagonally implicit Runge-Kutta} (DIRK) methods, for which ARKODE provides tables with orders of accuracy $q = \{2,3,4,5\}$ and embeddings of orders $p = \{1,2,3,4\}$, respectively.

As with ERKStep, the ARKStep module includes a number of features that allow users to optimize its usage on their problems. The full list of these is included in \cite{arkodeUserGuide}. One such feature is \editmade{the routine \texttt{ARKStepSetLinear},} that allows the user to specify that $f^I$ depends linearly on $y$, and thus each implicit stage equation for $z_i$ from \eqref{eq:ARK_method} corresponds to a linear system of equations. In this case, only one iteration of the Newton nonlinear solver is performed, avoiding extraneous nonlinear iterations and residual norm computations to assess convergence.

A second notable feature in ARKStep is its native interface to the scalable parallel-in-time (PinT) library XBraid \cite{xbraid-package}. Instead of performing a standard time-marching approach from one step to the next $(t_{n-1}\to t_{n})$, XBraid solves for all time values simultaneously using \emph{multigrid reduction in time} (MGRIT) \cite{falgout2014parallel}, a highly parallel iterative method that exposes parallelism in the time domain in addition to spatial parallelization. We have designed the ARKStep + XBraid interface so that simulation codes using ARKStep should require minimal modifications to explore using PinT.

\subsection{MRIStep}
\label{sec:mristep}

MRIStep is the newest time-stepping module in ARKODE and targets multirate IVPs of the form
\begin{equation}
  \label{eq:MRIStep_IVP_3comp}
  y'(t) = f^{E}(t,y) + f^{I}(t,y) + f^F(t,y), \qquad y(t_0) = y_0.
\end{equation}
The right-hand side function is split into slow components, $f^E(t,y) + f^I(t,y)$, that should be integrated with a large step size $H$, and where $f^E(t,y)$ contains nonstiff terms and $f^I(t,y)$ contains stiff terms; and fast components, $f^F(t,y)$, that should be integrated with a small step size $h \ll H$.

For approximating solutions to \eqref{eq:MRIStep_IVP_3comp}, MRIStep utilizes multirate infinitesimal (MRI) methods, that are characterized through an alternation between an outer ARK-based method for the slow time scale, and a sequence of fast time scale IVP systems that are evolved using an inner solver.

When both slow components $f^E$ and $f^I$ are present, MRIStep uses implicit-explicit multirate infinitesimal generalized-structure additive Runge--Kutta (ImEx-MRI-GARK) methods \cite{CR2021}.  When either $f^E$ or $f^I$ are \texttt{NULL}-valued, ImEx-MRI-GARK methods simplify to either implicit or explicit multirate infinitesimal GARK (MRI-GARK) methods \cite{sanduClassMultirateInfinitesimal2019}. The following algorithm outlines a single step $t_{n-1}\rightarrow t_{n-1}+H=t_n$ of an $s$-stage MRI method from either family:
\begin{enumerate}
\item Let $z_1 = y_{n-1}$ and $t_{n,1}=t_{n-1}$.
\item For $i = 2,\ldots,s$
\begin{enumerate}
\item Let $t_{n,i} = t_{n-1} + c_{i} H$,\; $\Delta c_i = c_i-c_{i-1}$,\; $v(t_{n,i-1}) = z_{i-1}$,  and \\
$r_i(t) = \frac{1}{\Delta c_i}\sum\limits_{j=1}^{i-1} \omega_{i,j}\left(\frac{t-t_{n,i-1}}{\Delta c_i H}\right) f^{E}(t_{n,j}, z_j) + \frac{1}{\Delta c_i}\sum\limits_{j=1}^i \gamma_{i,j}\left(\frac{t-t_{n,i-1}}{\Delta c_i H}\right) f^{I}(t_{n,j}, z_j)$.
\item Solve $v'(t) = f^F(t, v) + r_i(t)$ over $t \in [t_{n,i-1}, t_{n,i}]$. \label{step:fastIVP}
\item Set $z_i = v(t_{n,i})$.
\end{enumerate}
\item Set $y_{n} = z_{s}$.
\end{enumerate}
Here the coefficients, $0=c_1<\cdots<c_s= 1$, define the abscissae for the method.  The coefficient functions, $\omega_{i,j}$ and $\gamma_{i,j}$, are polynomials in time that dictate the couplings from the slow to the fast time scale and can be expressed as in \cite{CR2021} as
\begin{equation*}
  \omega_{i,j}(\theta) = \sum_{k=0}^K \omega_{i,j}^{\{k\}}\theta^k, \quad \gamma_{i,j}(\theta) = \sum_{k=0}^K \gamma_{i,j}^{\{k\}} \theta^k,
\end{equation*}
where typically $0\leq K \leq 2$. The coupling tables, $\Omega^{\{k\}}\in\mathbb{R}^{s\times s}$ and $\Gamma^{\{k\}}\in\mathbb{R}^{s\times s}$, contain the explicit, $\omega_{i,j}^{\{k\}}$, and implicit, $\gamma_{i,j}^{\{k\}}$, coefficients respectively.

Like Runge--Kutta methods, implicitness at the slow time scale is characterized by nonzero values on or above the diagonal. MRIStep currently supports diagonally implicit, ``solve-decoupled'' methods i.e., $\omega_{i,j}^{\{k\}} = 0$ for $j \geq i$, $\gamma_{i,j}^{\{k\}}=0$ for $j>i$, and $\Delta c_i=0$ if $\gamma_{i,i}^{\{k\}} \neq 0$. When $\Delta c_i=0$, the ``fast'' IVP solve in step \ref{step:fastIVP} reduces to a standard ImEx-ARK stage \eqref{eq:ARK_stage}, with step size, $H$, and coefficients
\begin{equation*}
  A^E_{i,j} = \sum_{k=0}^K \frac{\omega_{i,j}^{\{k\}}}{k+1}, \quad A^I_{i,j} = \sum_{k=0}^K \frac{\gamma_{i,j}^{\{k\}}}{k+1}.
\end{equation*}
Thus an MRI stage only requires a nonlinear implicit solve if $A^I_{i,i} \ne 0$.

MRIStep also supports Multirate infinitesimal step (MIS) methods \cite{SKAW2009,SKAW2012a,SKAW2012b}. These methods are a subset of MRI-GARK methods where the coupling coefficients are uniquely defined based on a slow explicit Butcher table $(A^E,b^E,c^E)$ with $s-1$ stages and sorted abscissae $c^E$. The MIS method then has abscissae, $c = [c^E_1\; \cdots\; c^E_{s-1}\; 1]^T$, and coupling coefficients, $\gamma_{i,j}^{\{0\}} = 0$, and
\begin{equation}
  \label{eq:MIS_to_MRI}
  \omega_{i,j}^{\{0\}} = \begin{cases}
    0, & \text{if}\; i=1,\\
    A^E_{i,j} - A^E_{i-1,j}, & \text{if}\; 2\le i\le s-1,\\
    b^E_{j} - A^E_{s-1,j}, & \text{if}\; i=s.
  \end{cases}
\end{equation}

At present, the MRIStep module implements all third and fourth order ImEx-MRI-GARK methods from \cite{CR2021}, the third-order MIS method defined by the slow Butcher table from \cite{KW1998}, as well as all of the third and fourth order MRI-GARK methods from \cite{sanduClassMultirateInfinitesimal2019}. Alternately, users may provide a custom set of coupling tables, $\Gamma^{\{k\}}$ and/or $\Omega^{\{k\}}$, to define their own MRI method.  MRIStep includes a utility routine that will also convert a slow Butcher table $(A^E,b^E,c^E)$ with sorted abscissae $c^E$ into a set of MRI-GARK coefficients using \eqref{eq:MIS_to_MRI}. If the slow Butcher table has order of accuracy at least two then the MIS method will also have order two, but third-order accuracy can be obtained if the slow table itself is at least third order and satisfies the condition \cite{KW1998}
\begin{equation*}
    \sum_{i=2}^{\hat{s}} \left(c_i^E-c_{i-1}^E\right) \left(e_i+e_{i-1}\right)^T A^E c^E
    + \left(1-c_{\hat{s}}^E\right) \left(\frac12+e_{\hat{s}}^T A^E c^E\right) = \frac13,
\end{equation*}
where $A^{E}$ has $\hat{s}$ stages and $e_k$ is the $k$-th column of an $\hat{s}\times \hat{s}$ identity matrix.

The ODEs in step \ref{step:fastIVP} may be solved using any sufficiently accurate method by supplying a ``fast IVP'' stepper to the MRIStep module.  For convenience, ARKODE includes a constructor that will attach an ARKStep instance as the inner integrator, thereby enabling an adaptive explicit, implicit, or ImEx treatment of the fast time scale. Alternatively, applications can supply a user-defined integrator for the fast time scale as demonstrated in the simple example problem \editmade{in Section \ref{sec:mristep_example}} and in the large-scale demonstration problem in Section \ref{sec:demo}.
Currently MRIStep supports temporal adaptivity only at the fast time scale. Efficient multirate temporal adaptivity is an open research question (see, e.g., \cite{FishReynolds2022}), but future support is anticipated to follow new developments in the field.  Finally, like ERKStep and ARKStep, MRIStep provides numerous configuration options for users to control integrator behavior.  The full list of options is included in \cite{arkodeUserGuide}.

\section{Shared Infrastructure}
\label{sec:infrastructure}

As mentioned above, the ARKODE time-stepping modules rest on a shared infrastructure providing various utilities for one-step methods.  This infrastructure manages the time loop, calling the time stepper modules both to advance the solution with a given step size and to compute a temporal error estimate for that step.  This outer loop handles the temporal error control and time step adaptivity, the detection of events (root finding) and solution constraint violations, and, if necessary, interpolation of the solution to a requested output time. Moreover, ARKODE provides time steppers with interfaces to algebraic solvers and methods for predicting the new time solution.  We note that ARKODE itself rests on the shared SUNDIALS infrastructure.  Thus, throughout this section we focus on the ARKODE-specific utilities, deferring to more general SUNDIALS references \cite{HBGLSSW2005,gardner2020enabling,balosEnablingGPUAccelerated2021} for descriptions of that shared infrastructure.

\subsection{User-supplied Tolerances and Error Control}
\label{sec:tolerances_error_control}

To control errors at various levels (time integration and algebraic solvers), ARKODE, like other SUNDIALS solvers \cite[eq.~(4)]{HBGLSSW2005}, uses a \emph{weighted root-mean-square norm} for all error-like quantities,
\begin{equation}
  \label{eq:wrms_norm}
  \|v\|_{\text{WRMS}} = \left( \frac{1}{N} \sum_{i=1}^N \left(v_i\,w_i\right)^2\right)^{1/2}.
\end{equation}
This norm allows problem-specific error measurement via the weighting vector, $w$, that combines the units of the problem with user-supplied values that specify ``acceptable'' error levels.  Thus, at the start of each time step $t_{n-1} \to t_n$, we construct two weight vectors.  The first is an \emph{error weight vector} that, like in CVODE \cite[eq.~(5)]{HBGLSSW2005}, uses the most recent step solution and user-supplied relative and absolute tolerances (scalar- or vector-valued),
\begin{equation}
  \label{eq:ewt}
  w_i = \left(\text{rtol}\, |y_{n-1,i}| + \text{atol}_{i}\right)^{-1}.
\end{equation}
The second focuses on problems with a non-identity \editmade{operator $M(t)$}, since the units of \eqref{eq:IVP} may differ from the units of the solution $y$.  Here, we construct a \emph{residual weight vector},
\begin{equation}
  \label{eq:rwt}
  \editmade{\sigma_i} = \left(\text{rtol}\, |r_i| + \text{ratol}_i\right)^{-1},  \quad r = M(t_{n-1}) y_{n-1},
\end{equation}
where the user may specify a separate residual absolute tolerance value or array, $\text{ratol}$.

ARKODE uses $w$ or \editmade{$\sigma$} based on the quantity being measured: errors having ``solution'' units use $w$, whereas those having ``equation'' units use \editmade{$\sigma$}.  Obviously, when $M=I$ the solution and equation units match, and ARKODE uses $w$ for all error norms.  In both cases, since $w_i^{-1}$ (or \editmade{$\sigma_i^{-1}$}) represents a tolerance in the $i$-th component of the solution (or equation) vector, then an error whose WRMS norm is $\le 1$ is regarded as being within an acceptable error tolerance.

\subsection{Stepsize Adaptivity}
\label{sec:time_adaptivity}

A critical utility provided by ARKODE is its adaptive control of local truncation error (LTE), $T_n$, at each step, which corresponds to the error induced within the time step $t_{n-1}\to t_n$, under an assumption that the initial condition for that step was \emph{exact}.  At every internal step, ARKODE requests both an approximate solution $y_n$, and an estimate of $T_n$, from the time-stepping module.  These  approximations have global orders of accuracy, $q$ and $p$, respectively, where generally $p=q-1$.  Since the global and local errors for one step methods usually differ by one factor of $h_n$ \cite{HNW2000}, then
\begin{align}
  \label{eq:SolutionError}
  \| y_n - y(t_n) \| = C h_n^{q+1} + \bigO{h_n^{q+2}}, \\
  \label{eq:ErrorEstimate}
  \| T_n \| = D h_n^{p+1} + \bigO{h_n^{p+2}}.
\end{align}
where $C$ and $D$ are constants independent of $h_n$, and where we have assumed exact initial conditions for the step, i.e. $y_{n-1} = y(t_{n-1})$.  Given this time-stepper-estimated value of $T_n$, ARKODE adopts the same local temporal error test as other SUNDIALS integrators, simply
\begin{equation}
  \label{eq:temporal_error_test}
  \|T_n\|_{\text{WRMS}} \le 1.
\end{equation}
If this test passes, the step is accepted and $T_n$ is used to estimate a prospective next step size, $h'$, using an adaptive error controller.  If \eqref{eq:temporal_error_test} fails then the step is rejected, and $T_n$ is used to estimate a reduced step size, $h'$.  In either case, temporal adaptivity in ARKODE focuses on choosing the maximum $h'$ such that \eqref{eq:temporal_error_test} should pass on the next attempted step.  We thus define the \emph{biased} local error estimate
\[
  \varepsilon_k \ \equiv \ \beta\, \|T_k\|_{\text{WRMS}},
\]
where the error bias \editmade{$\beta>1$} helps to account for the error constant $D$ in \eqref{eq:ErrorEstimate}.  ARKODE stores the biased error estimates corresponding to the three most recent steps, $\varepsilon_n$, $\varepsilon_{n-1}$ and $\varepsilon_{n-2}$.  Of these, both $\varepsilon_{n-1}$ and $\varepsilon_{n-2}$ correspond to the last two \emph{successful} steps, whereas $\varepsilon_n$ corresponds to the most-recently \emph{attempted} step.  ARKODE provides a variety of error controllers that utilize these estimates, including \emph{PID} (the ARKODE default) \cite{KC2003,S1998,S2003,S2006}, \emph{PI} \cite{KC2003,S1998,S2003,S2006}, \emph{I} (the ``standard'' method used in most publicly-available IVP solvers) \cite{HS2012}, the \emph{explicit Gustafsson} controller \cite{G1991}, the \emph{implicit Gustafsson} controller \cite{G1994}, and an ``ImEx'' Gustaffson controller that sets $h'$ as the minimum of the explicit and implicit Gustafsson controller values.  Alternately, users may provide their own functions to produce $h'$ using the inputs $y_n, t_n, h_n, h_{n-1}, h_{n-2}, \varepsilon_n, \varepsilon_{n-1}$, and $\varepsilon_{n-2}$.  Each controller bases its adaptivity off the order of accuracy for the error estimate, $p$, although the user may instead request to use the solution order $q$ instead.  All of the provided controllers include tuning parameters, with default values determined using an exhaustive genetic optimization process based off of dozens of challenging IVP test problems, including nearly all from the ``Test Set For IVP Solvers'' \cite{IVPTestSet}.

In addition to error-based temporal adaptivity, ARKODE adopts many of the heuristics from CVODE \cite{HBGLSSW2005} for bounding $h'$, including approaches for handling repeated temporal error test failures, overly aggressive growth or decay in prospective steps ($h' \gg h_n$ or $h' \ll h_n$), step size reduction following a failed implicit solve, and user-supplied step size bounds  $h_\text{min} \le h' \le h_\text{max}$.  Complete details on these heuristics, as well as information on how each parameter may be adjusted by users, are provided in the ARKODE documentation \cite{arkodeUserGuide}.

\subsection{Interpolation Modules}
\label{sec:interpolation}

ARKODE supports interpolation of approximate solutions $y(t_{out})$, and derivatives $y^{(d)}(t_{out})$, where $t_{out}$ occurs within the most recently completed step $t_{n-1}\to t_n$.  This utility also supports extrapolation of $y$ and its derivatives outside this time step to construct predictors for iterative algebraic solvers.  These approximations are based on polynomial interpolants $\pi_{\xi}(t)$ of degree up to $\xi=5$, that are constructed using one of two complementary approaches: ``Hermite'' and ``Lagrange''.

\subsubsection{Hermite interpolation module}
\label{sec:interpolation_hermite}

For non-stiff problems ARKODE provides Hermite polynomial interpolants, constructed using both solution and derivative data ($y_n \approx y(t_n)$ and $f_n \equiv f(t_n,y_n)\approx y'(t_n)$), following the approach outlined in \cite[II.6]{HNW2000}.  The available Hermite interpolants are:
\begin{itemize}
\item $\pi_0(t)$: satisfies $\pi_0(t) = \tfrac12\left(y_{n-1} + y_{n}\right)$.
\item $\pi_1(t)$: satisfies $\pi_1(t_{n-1}) = y_{n-1}$ and $\pi_1(t_n) = y_{n}$.
\item $\pi_2(t)$: satisfies $\pi_2(t_{n-1}) = y_{n-1}$,\: $\pi_2(t_n) = y_n$,\: and $\pi_2'(t_n) = f_{n}$.
\item $\pi_3(t)$: satisfies $\pi_3(t_{n-1}) = y_{n-1}$,\: $\pi_3(t_n) = y_n$,\: $\pi_3'(t_{n-1}) = f_{n-1}$,\: and $\pi_3'(t_n) = f_{n}$.
\item $\pi_4(t)$: satisfies $\pi_4(t_{n-1}) = y_{n-1}$,\: $\pi_4(t_{n}) = y_{n}$,\: $\pi_4'(t_{n-1}) = f_{n-1}$,\: $\pi_4'(t_n) = f_{n}$,\: and\\ $\pi_4'\left(t_{n} - \tfrac{h_n}{3}\right) = f\left(t_{n} - \tfrac{h_n}{3},\, \pi_3\left(t_{n} - \tfrac{h_n}{3}\right)\right)$.
\item $\pi_5(t)$: satisfies $\pi_5(t_{n-1}) = y_{n-1}$,\: $\pi_5(t_{n}) = y_{n}$, $\pi_5'(t_{n-1}) = f_{n-1}$,\: $\pi_5'(t_n) = f_{n}$,\:\\ $\pi_5'\left(t_{n} - \tfrac{h_n}{3}\right) = f\left(t_{n} - \tfrac{h_n}{3},\, \pi_4\left(t_{n} - \tfrac{h_n}{3}\right)\right)$,\: and
$\pi_5'\left(t_{n} - \tfrac{2h_n}{3}\right) = f\left(t_{n} - \tfrac{2h_n}{3},\, \pi_4\left(t_{n} - \tfrac{2h_n}{3}\right)\right)$.
\end{itemize}
The interpolants $\pi_4(t)$ and $\pi_5(t)$ require one and four additional evaluations of $f(t,y)$, respectively, whereas all others are constructed using readily-available data from the last successful time step.  Due to these increasing costs, interpolants of degree $\xi > 5$ are not currently provided.

\subsubsection{Lagrange interpolation module}
\label{sec:interpolation_lagrange}

For stiff problems, interpolants that use derivative values $f=y'$ can result in order reduction.  We thus provide a second module based on polynomial interpolants of Lagrange type, constructed using an extended ``history'' of solution data, $\left\{ y_{n}, y_{n-1}, \ldots, y_{n-\xi} \right\}$,
\begin{align*}
  \pi_{\xi}(t) = \sum_{j=0}^{\xi} y_{n-j}\, \ell_j(t),\quad\text{where}\quad
  \ell_j(t) = \prod_{\substack{i=0\\ i\ne j}}^{\xi} \left(\frac{t-t_i}{t_j-t_i}\right), \; j=0,\ldots,\xi.
\end{align*}
Since generally each solution vector $y_{n-j}$ has many more than 5 entries, we evaluate $\pi_{\xi}$ at any desired $t\in\R$ by first evaluating the basis functions at $t$, and then performing a linear combination of the stored solution vectors $\{y_{n-k}\}_{k=0}^{\xi}$.  Derivatives $\pi_{\xi}^{(d)}(t)$ may be evaluated similarly as
\[
  \pi_{\xi}^{(d)}(t) = \sum_{j=0}^{\xi} y_{n-j}\, \ell_j^{(d)}(t).
\]
Due to the increasing algorithmic complexity involved in evaluating $\ell_j^{(d)}$, ARKODE only supports derivatives up to $d=3$.  Also, since in the first $(\xi-1)$ internal time steps ARKODE has an insufficient solution history to construct the full $\xi$-degree interpolant, during these initial steps we construct the highest degree interpolants that are currently available.

\subsection{Implicit Solvers}
\label{sec:implicit_solvers}

For both ARKStep and MRIStep, if $f^{I}$ is nonzero then each implicit stage, $z_i\in\R^N$, may require the solution of an algebraic system of equations:
\begin{equation}
  \label{eq:implicit}
  M(t_{n,i})\left(z_i - a_i\right) - \gamma_i f^I(t_{n,i},z_i) = 0,
\end{equation}
where $t_{n,i}$ is the corresponding implicit stage time, $\gamma_i\in\R$ is proportional to $h_n$, $f^I(t,y)$ is the implicit portion of the IVP right-hand side, and $a_i$ is ``known data'' that may include previous stages or IVP right-hand side evaluations.  ARKODE provides utilities to map between $f^I$ and the nonlinear or linear algebraic system for each stage \eqref{eq:implicit},  interfaces to a range of algebraic solvers from SUNDIALS, and optimizations to enhance the efficiency of these solvers for the IVP context.

\subsubsection{Nonlinear Solver Methods}
\label{sec:nonlinear_solvers}

SUNDIALS provides multiple nonlinear solver modules through its \texttt{SUNNonlinearSolver} interface \cite{gardner2020enabling}, all of which utilize predictor-corrector form and are targeted to nonlinear systems of equations with either root-finding or fixed-point structure.  Writing the stage solution as $z_i=z_{i,p}+z_{i,c}$, where $z_{i,p}$ is the prediction and $z_{i,c}$ the desired correction, ARKODE rewrites the nonlinear system \eqref{eq:implicit} in root-finding or fixed-point form as
\begin{subequations}
\label{eq:predictor_corrector}
\begin{align}
  \label{eq:predictor_corrector_rf}
  0 &= G_{rf}(z_{i,c}) := M(t_{n,i})z_{i,c} - \gamma_i f^I(t_{n,i},z_{i,p}+z_{i,c}) - M(t_{n,i})\left(a_i-z_{i,p}\right),\quad\text{or}\\
  \label{eq:predictor_corrector_fp}
  z_{i,c} &= G_{fp}(z_{i,c}) := M(t_{n,i})^{-1} \left(\gamma_i f^I(t_{n,i},z_{i,p}+z_{i,c}) \right) + (a_i - z_{i,p}),
\end{align}
\end{subequations}
respectively. Note, these equations may be simplified in the cases where $M$ is independent of $t$ or when $M=I$.
When setting up an ARKODE simulation, users must identify the category of $M$ at problem initialization, at which point ARKODE will provide a function pointer for the relevant nonlinear system to the generic \texttt{SUNNonlinearSolver} object.  This ARKODE layer serves to translate from the user-supplied functions $M$ and $f^I$ and user-selected IVP method to the generic nonlinear solver module.  To simplify the discussion below, we focus on the general cases \eqref{eq:predictor_corrector}.

Nonlinear solvers for root-finding problems typically solve linear systems related to the Jacobian,
\begin{equation}
\label{eq:Jacobian_definition}
   \mathcal{A}(t,z,\gamma) := \frac{\partial G_{rf}}{\partial z_{i,c}} = M(t) - \gamma \frac{\partial f^I(t,z)}{\partial z}.
\end{equation}
When either the matrix $\mathcal{A}$ is explicitly stored or a preconditioner is supplied by the user, ARKODE updates these infrequently to amortize the high costs of Jacobian construction and preconditioner formulation. Thus, ARKODE will use a Jacobian or preconditioner corresponding to \eqref{eq:Jacobian_definition} at a previous time step i.e., $\tilde{\mathcal{A}}(\tilde{t},\tilde{z},\tilde{\gamma})$ where $\tilde{z}$, $\tilde{t}$ and $\tilde{\gamma}$ are previous time step values. The logic guiding this lagging process follows the heuristics described for CVODE in \cite[Sec.~2.1]{HBGLSSW2005} and each of the specific heuristic parameters may be modified by the user for their problem.

ARKODE utilizes an identical stopping test for the nonlinear iteration as is used in CVODE, that strives to solve the nonlinear system slightly more accurately than the temporal accuracy in an effort to minimize extraneous (and costly) nonlinear solver iterations. As with the other heuristics in ARKODE all of the stopping test parameters are user modifiable.

\subsubsection{Implicit Predictors}
\label{sec:implicit_predictors}

For iterative nonlinear solvers, a good initial guess can dramatically affect both their speed and robustness, making the difference between rapid quadratic convergence versus divergence of the iteration.  To this end, ARKODE provides a variety of algorithms to construct the predictor $z_{i,p}$ from \eqref{eq:predictor_corrector}, typically using an interpolating polynomial via ARKODE's interpolation module (Section \ref{sec:interpolation}).  Specifically, since each stage solution typically satisfies $z_i \approx y(t_{n,i})$, where $t_{n,i}$ is the stage time associated with $z_i$, then we predict these stage solutions as
\begin{equation}
  \label{eq:extrapolant}
  z_{i,p} = \pi_\xi(t_{n,i}).
\end{equation}
Since $t_{n,i}$ are usually outside of the previous successful step, $[t_{n-2},t_{n-1}]$ (containing the data used to construct $\pi_\xi(t)$), \eqref{eq:extrapolant} will correspond to an extrapolant instead of an interpolant.  The dangers of using polynomial extrapolation are well-known, with higher-order polynomials and evaluation points further outside the interpolation interval resulting in the greatest risk.  To support ``optimal'' choices for different types of problems, ARKODE's prediction algorithms construct a variety of interpolants with different degree and using different interpolation data, as described below.

\textbf{Trivial predictor}.
This \editmade{predictor} is given by $\pi_0(t) = y_{n-1}$.  While not highly accurate, this predictor is often the most robust choice for very stiff problems or for problems with implicit constraints whose violation may cause illegal solution values (e.g., a negative density or temperature).

\textbf{Maximum order predictor}.
At the opposite end of the spectrum, we may construct the highest-degree interpolant available, $\pi_{\xi_\text{max}}(t)$.  We note that as $\xi_\text{max}$ increases, this predictor provides a very accurate prediction for stage times that are ``close'' to $[t_{n-2},t_{n-1}]$ but could be dangerous otherwise.  As discussed in Section \ref{sec:interpolation}, ARKODE caps the polynomial degree at 5, but the degree may be further limited by setting, $\xi_\text{user}$, and the IVP solver method order, $q$, such that $\xi_\text{max} \le \min\{q-1,\xi_\text{user},5\}$.

\textbf{Variable order predictor}.
In-between the two previous approaches, this predictor uses $\pi_{\xi_\text{max}}(t)$ for predicting earlier stages and lower-degree polynomials for later stages.  Thus, the polynomial degree is chosen adaptively based on the stage index $i$, $\xi = \max\{ \xi_\text{max} - i,\; 1 \}$, which may be reasonable under the assumption that the stage times are increasing, i.e., $c_i \le c_j$ for $i<j$.

\textbf{Cutoff order predictor}.
Following a similar idea as above, this predictor monitors the actual stage times to determine the polynomial degree for $\pi_\xi(t)$:
\begin{equation*}
  \xi = \begin{cases}
    \xi_\text{max}, & \text{if}\quad \frac{t_{n,i}-t_{n-1}}{h_{n-1}} < \tfrac12,\\
    1, & \text{otherwise}.
\end{cases}
\end{equation*}

\textbf{User supplied}.
Finally, ARKODE supports a user-provided implicit predictor function that is called after the selected built-in routine. Thus it could predict all components of $z_{i,p}$ if desired, or it could merely ``fix'' specific components that result from a higher order built-in predictor (e.g., to satisfy underlying physical constraints) before the prediction is passed to the nonlinear solver.

\subsubsection{Linear Solver Methods}
\label{sec:linear_solvers}

For problems that require linear solves within a nonlinear solver or the action of $M(t)^{-1}$, users may leverage any of the \texttt{SUNLinearSolver} and \texttt{SUNMatrix} modules provided by SUNDIALS \cite{gardner2020enabling}.  These solvers generally fit within one of four categories: (a) direct solvers that store and operate on matrices, (b) iterative and matrix-based solvers, (c) iterative and matrix-free solvers, or (d) solvers that embed the linear system.  For linear solvers of each type, ARKODE provides utilities to assist in translating from the user-supplied, problem-defining functions, $f^I$, $M(t)$ (or the product, $M(t)\,v$), and, optionally, the Jacobian, $J(t,z) = \partial_z f^I(t,z)$, or product, $J(t,z)\,v$, to the generic SUNDIALS modules.  In the case of matrix-based linear solvers that utilize dense or banded matrices, ARKODE provides utilities to approximate the Jacobian matrix, $J$, using a minimum number of evaluations of $f^I$ \cite{HBGLSSW2005}.

As with the nonlinear case, ARKODE attempts to solve each linear system to an accuracy just below that of the encompassing routine (e.g., an outer nonlinear solver in the case of $\mathcal{A}$ from \eqref{eq:Jacobian_definition}, or the temporal error controller in the case of $M^{-1}$).  Again, ARKODE follows the same techniques for this purpose as CVODE \cite{HBGLSSW2005}, with the only salient differences being ARKODE's use of the residual weight vector, \editmade{$\sigma$}, from \eqref{eq:rwt} for linear system residual norms. As before, all of the associated heuristic parameters are user modifiable.
When solving problems with both a non-identity mass matrix and a matrix-based nonlinear solver, the ``types'' of the mass matrix and system \texttt{SUNLinearSolver} objects must match (matrix-based, matrix-based iterative, matrix-free iterative, or embedded). If both are matrix-based, these matrices must use the same \texttt{SUNMatrix} storage type for streamlined construction of the combination, $M - \gamma J$.
Otherwise, ARKODE has no restriction that the solvers themselves be the same (e.g., a basic conjugate gradient method could be used for the mass matrix systems, along with a preconditioned generalized minimum residual method for the system matrix $\mathcal{A}$).

\subsection{Temporal root-finding}
\label{sec:rootfinding}

A particularly useful feature of SUNDIALS integrators and provided in the ARKODE infrastructure is event detection.  While integrating an IVP, ARKODE can find roots of a set of user-defined functions, $g_i(t,y(t)) = \tilde{g}_i(t)$. The number of these functions is arbitrary, and, if more than one $\tilde{g}_i$ is found to have a root in any given interval, the various root locations are found and reported in the order that they occur in the direction of integration.
The basic scheme and heuristics match what was used elsewhere in SUNDIALS \cite{HBGLSSW2005}.  During integration of \eqref{eq:IVP}, following each successful step ARKODE checks for sign changes of any $\tilde{g}_i(t)$; if one is found, it then will home in on the root (or roots) with a modified secant method \cite{HS1980}.

\subsection{Inequality Constraints}
\label{sec:constraints}

A final notable feature of the ARKODE infrastructure, also shared with other SUNDIALS integrators, is support for user-imposed inequality constraints on components of the solution $y$: $y_i > 0$, $y_i < 0$, $y_i \geq 0$, or $y_i \leq 0$. Constraint satisfaction is tested after a successful step and before the temporal error test. If any constraint fails, ARKODE estimates a reduced step size based on a linear approximation in time for each violated component (including a safety factor to cover the strict inequality case). If applicable, a flag is set to update the Jacobian or preconditioner in the next step attempt.

\section{Usage}
\label{sec:usage}

Each time stepping module in ARKODE admits a similar usage style, modeled after CVODE \cite[Sec.~6]{HBGLSSW2005}.  Moreover, due to the recent addition of Fortran 2003 interfaces throughout SUNDIALS \cite{gardner2020enabling}, users may call ARKODE from C, C++, or Fortran by following the same progression of steps. This general approach is as follows:
\begin{enumerate}
  \item The user provides one or more functions that define the IVP to be solved, \editmade{e.g.}, $f(t,y)$, $f^I(t,y)$, as well as any problem-specific preconditioners or utility routines.
  \item For problems with a non-identity mass matrix, the user provides a function to either compute the mass matrix, $M(t)$, or to perform the mass matrix-vector product, $M(t)\, v$.
  \item The user creates a vector of initial conditions, $y_0$, for the problem \eqref{eq:IVP}.
  \item The user creates the ARKODE time-stepper memory structure that contains default values for solver options, such as solution error tolerances and time-stepping method order.
  \item The user creates nonlinear and/or linear solver objects for use by the stepper, e.g., a Newton solver, a sparse linear solver, and a sparse matrix.  These objects also include default options.
  \item The user attaches the algebraic solver objects to the ARKODE time-stepper.
  \item \label{skel:user_set}
  The user adjusts time-stepper or algebraic solver options via ``set'' routines.  The user may additionally specify temporal root-finding functions, inequality constraints, etc.
  \item The user advances the solution toward a desired output time, $t_{out}$, using one of four ``modes'':
  \begin{itemize}
    \item \emph{NORMAL} takes internal steps until simulated time has just passed $t_{out}$ and returns an approximation of $y(t_{out})$ using interpolation (see Section \ref{sec:interpolation}).
    \item \emph{ONE-STEP} takes a single internal step and returns to the user. If this step overtakes $t_{out}$, then the solver approximates $y(t_{out})$; otherwise it returns $y_n$.
    \item \emph{NORMAL-TSTOP} takes internal steps until the next step will overtake $t_{out}$. The subsequent step is limited so that $t_n = t_{out}$, and the internal solution $y_n$ is returned.
    \item \emph{ONE-STEP-TSTOP} takes a single internal step and returns to the user (like ``one-step'').  However, if the step will overtake $t_{out}$, then this mode is identical to ``normal-tstop''.
  \end{itemize}
  Most users call ARKODE in either NORMAL or NORMAL-TSTOP mode, where the latter is preferred when full integrator accuracy is required (to avoid interpolation error).  The two ONE-STEP modes are frequently used when debugging or first beginning with ARKODE, in order to verify the results following each internal step.
  \item The user retrieves optional outputs from the time-stepper via ``get'' routines, including solver statistics or interpolated solution values.
  \item The user may repeat the above process starting at step \ref{skel:user_set}.
  \item When the solution process is complete, the user destroys any SUNDIALS objects they have created e.g., solution vectors, algebraic solvers, time integrators, etc.
\end{enumerate}

A variety of additional options are supported within this basic structure.  Three particularly useful ARKODE features include the ability to \emph{re-initialize}, \emph{reset}, and \emph{resize} the integrator.
Re-initialization is useful when a user has already solved an IVP using ARKODE and wants to solve a new problem with a state vector of the same size.  Here, no memory is (de)allocated, but all solver history (previous step sizes, temporal error estimates, stored solutions, solver statistics) is cleared.  This is useful when performing parameter sweeps, where the user wishes to run repeated forward simulations using different problem parameters.  This is also useful when treating jump discontinuities in the IVP right-hand side functions that cause solution characteristics to change dramatically.  Here, one may run ARKODE up to the discontinuity in ``tstop'' or root-finding mode, and then re-initialize the solver with the new right-hand side function to be used following the discontinuity.
Alternately, ARKODE steppers may be ``reset'', allowing them to solve an IVP with the same right-hand side function(s), but using an updated initial condition $(t_0,y_0)$.  As with re-initialization, the memory footprint remains unchanged and the step size and solution history are cleared, but all solver statistics (total number of steps, etc.) are retained.  This mode is used by MRIStep to ensure that each fast time scale IVP is solved using the prescribed initial condition, while still accumulating all fast time scale solver statistics.
Finally, since ARKODE is built for one-step methods, it is natural to apply it to PDEs that adaptively refine the spatial grid.  Here, the ARKODE ``resize'' functions support simulations in which the number of equations and unknowns in the IVP system change between time steps.  This resizing obviously modifies ARKODE's internal memory structures to use the new problem size, but does so without destroying the temporal adaptivity heuristics.

\section{Numerical Results}
\label{sec:demonstration}

To highlight the algorithmic flexibility provided by ARKODE, we first present a sequence of simple numerical examples that progressively increase in both problem and integrator complexity, emulating the progression of a typical new ARKODE user.  We then present a demonstration problem showing an advanced use of ARKODE for a large-scale multiphysics application.

\subsection{Simple Example Problem}
\label{sec:simple-example}

For each stepper module example, we consider a one-dimensional advection-diffusion-reaction equation using a variation of the Brusselator chemical kinetics model \cite{HW2002}. The system is given by
\begin{equation} \label{eq:ex-adv-diff-react}
\begin{aligned}
    u_t &= -c\, u_x + d\, u_{xx} + a - (w + 1)\, u + v\, u^2, \\
    v_t &= -c\, v_x + d\, u_{xx} + w\, u - v\, u^2, \\
    w_t &= -c\, w_x + d\, u_{xx} + (b - w) / \epsilon - w\, u,
\end{aligned}
\end{equation}
for $t \in [0, 10]$ with $x \in [0, 1]$, where $u$, $v$, and $w$ are the chemical species concentrations, $c = 0.001$ is the advection speed, $d$ is the diffusion rate (typically $0.01$), $a = 0.6$ and $b = 2$ are the concentrations of species that remain constant over time, and $\epsilon = 0.01$ is a parameter that determines the stiffness of the system. The initial conditions are
\begin{equation*}
\begin{gathered}
    u(0,x) =  a  + 0.1 \sin(\pi x), \qquad
    v(0,x) = b/a + 0.1 \sin(\pi x), \qquad
    w(0,x) =  b  + 0.1 \sin(\pi x).
\end{gathered}
\end{equation*}
Spatial derivatives are computed using second-order centered differences with the data distributed over $N = 512$ points on a uniform spatial grid with stationary boundary conditions i.e.,
\begin{equation*}
\begin{gathered}
    u_t(t,0) = u_t(t,1) = 0, \qquad
    v_t(t,0) = v_t(t,1) = 0, \qquad
    w_t(t,0) = w_t(t,1) = 0.
\end{gathered}
\end{equation*}
\editmade{In the subsections that follow, we explore various time integration methods and ImEx splittings of \eqref{eq:ex-adv-diff-react}, so we first summarize the relevant time scales present in this problem for each of the advection, diffusion, and reaction terms.  For this problem setup, if explicit time stepping methods were used for all operators and linear stability were the only consideration, then diffusion would require the smallest time step, followed by reaction, followed by advection.  However, if linear stability were not an issue (e.g., when using $A$-stable implicit methods) and temporal accuracy were the only consideration, then reaction would require the smallest time steps, followed by advection and diffusion that would evolve on comparable time scales.  Finally, we note that the $(b-w)/\epsilon$ term in \eqref{eq:ex-adv-diff-react} induces a mild amount of additional stiffness into the reaction processes, thus requiring slightly smaller ``stable'' than ``accurate'' steps.}

All of the examples use the serial N\_Vector implementation and results were obtained on \editmade{a single node of} the Quartz cluster at Lawrence Livermore National Laboratory. \editmade{Each Quartz node contains two Intel Xeon 5-2695 v4 CPUs.} The source files for this example problem are
\href{https://github.com/LLNL/sundials/tree/v6.5.0/examples/arkode/CXX_serial/ark_advection_diffusion_reaction.hpp}{ark\_advection\_diffusion\_reaction.hpp}
and
\href{https://github.com/LLNL/sundials/tree/v6.5.0/examples/arkode/CXX_serial/ark_advection_diffusion_reaction.cpp}{ark\_advection\_diffusion\_reaction.cpp}. \editmade{In each test case the maximum relative error at the final time is computed using a reference solution generated with the default fifth-order DIRK method (ARK5(4)8L[2]SA-ESDIRK in \cite{KC2003}) with relative and absolute tolerances of $10^{-8}$ and $10^{-14}$, respectively.}

\subsubsection{ERKStep}
\label{sec:erkstep_example}

We first consider an advection-reaction setup (i.e., $d = 0$ in \eqref{eq:ex-adv-diff-react}) and evolve the system in time using the default second- (Heun-Euler), third- \cite{BS1989}, fourth- \cite{Z1963}, and fifth-order \cite{CK1990} ERK methods in ARKODE. The optimal method choice will depend on the user's desired accuracy, the cost per step of the method, and the step sizes selected by the adaptivity controller. As such, we show how the options in ARKODE can help users determine more effective methods in their accuracy regimes of interest.  We compare the performance of each method using ``loose,'' ``medium,'' and ``tight'' sets of relative and absolute tolerances pairs ($10^{-4}$/$10^{-9}$, $10^{-5}$/$10^{-10}$, and $10^{-6}$/$10^{-11}$, respectively) and the PID, PI, I, and explicit Gustafsson \editmade{controllers.}

\begin{figure}[htb!]
    \centering
    \includegraphics[width=0.95\textwidth]{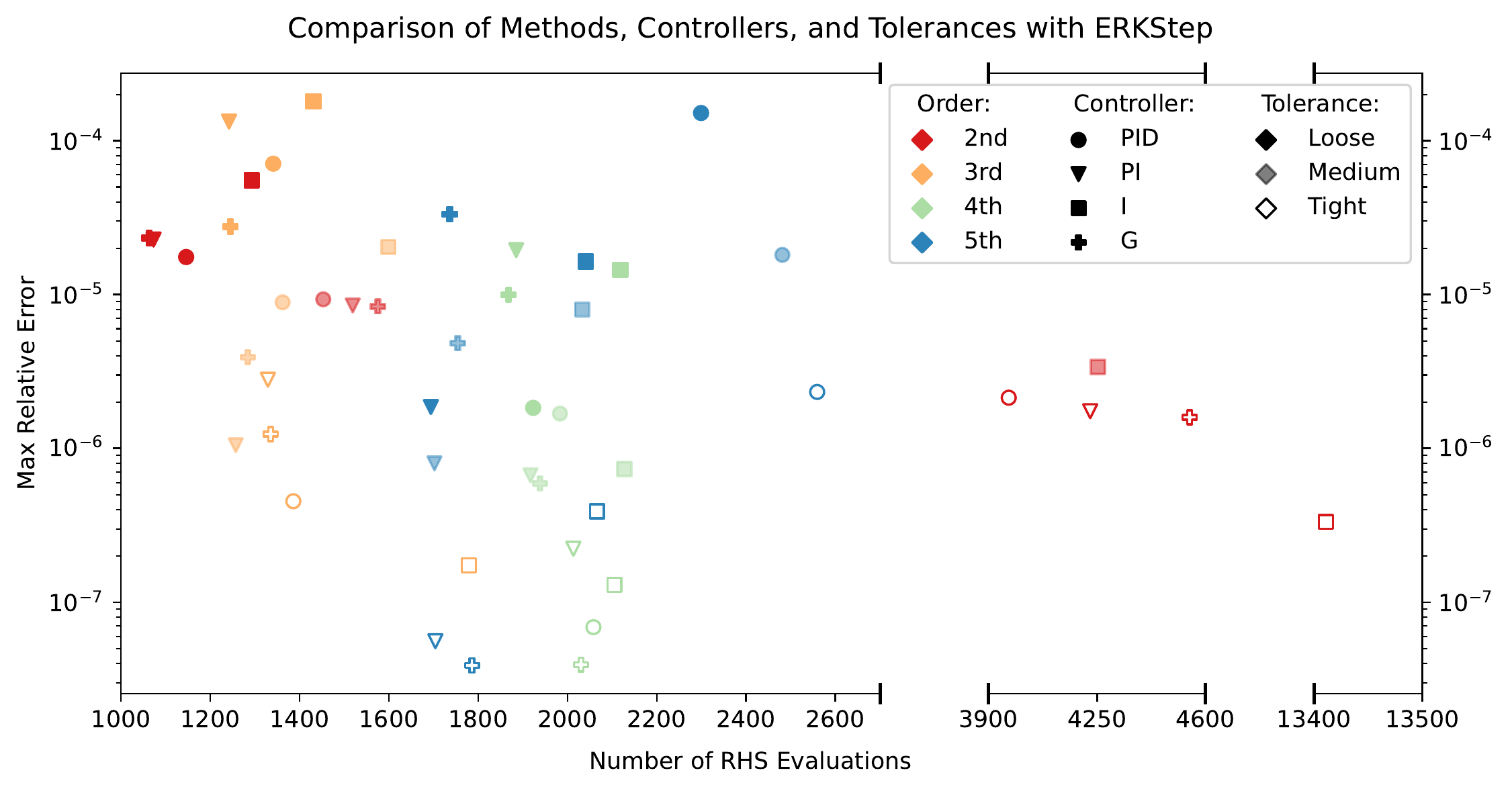}
    \caption{Work-precision plot for the default 2nd to 5th order ERK methods with various error controllers and tolerances. Configurations further to the left and lower in the plot indicate methods with less work and greater accuracy, respectively.}
    \label{fig:erk-results}
\end{figure}

Fig. \ref{fig:erk-results} shows a work-precision plot for the various configurations \editmade{where the total number of right-hand side function evaluations for successful and failed steps is used as the work metric}. Given its lower cost per step, the second-order method gives the best performance at coarser levels of accuracy (three left-most solid red markers) despite requiring more than double the number of step attempts as the third-order method. For errors between $10^{-5}$ and $10^{-7}$ the third-order method is the most efficient (lower left orange markers) since the second-order method requires at least three times as many step attempts, while the fourth- and fifth-order methods make fewer attempts but the reduced number of attempts is insufficient to offset their higher cost per step. The fifth-order method becomes most efficient at accuracies below $10^{-7}$ (blue markers near the bottom left).

Across the methods and tolerances, the I controller is frequently the least efficient for this example as it only considers the error in the current step leading to a high step rejection rate (between 13\% and 50\%). The additional error history retained by the PI and Gustafsson controllers leads to better step size predictions and often the greatest efficiency (rejection rates  <7\%). The longer history of the PID controller does not lead to further improvement and it generally produces results that fall between the PI and I controllers. As explored in \cite{ranocha2021optimized}, the choice of controller parameters (tunable in ARKODE but not explored here) also plays an important role in performance.

\subsubsection{ARKStep}
\label{sec:arkstep_example}

We extend the previous model to now include diffusion (i.e., $d=0.01$ in \eqref{eq:ex-adv-diff-react}).  To avoid the stability limitations of a fully-explicit approach, users could either treat the full system implicitly with a DIRK method or partition the right-hand side and apply an ImEx-ARK method with implicit diffusion. In the latter case, different splitting approaches could be considered, and in both cases the choice of predictor may not be clear.  As such, we show how these methods could be compared. In the DIRK case, we advance the system in time using the implicit part of the ImEx-ARK method, ARK4(3)6L[2]SA, from \cite{KC2003}, and, in the ImEx case, we use the ARK4(3)6L[2]SA ImEx-ARK method.  We consider two different ImEx splittings that both treat advection explicitly\editmade{, as is typical when evolving hyperbolic terms (particularly for flows with discontinuities)}, but the reaction terms are either treated implicitly (IMEX-1) to step over the fastest reaction dynamics or explicitly (IMEX-2) to track the reaction timescale. In each setup, we solve the nonlinear systems at each implicit stage using a modified Newton method paired with a banded direct linear solver. Additionally, in the ImEx-2 case we enable the ``linearly-implicit'' ARKStep option to allow only one Newton iteration per nonlinear solve. The integration relative and absolute tolerances are set to $10^{-4}$ and $10^{-9}$, respectively. For each method we compare the trivial, max order, variable order, and cutoff implicit predictor algorithms.

\begin{table}[htb!]
\centering
\caption{Comparison of integrator statistics using the default 4th order DIRK and ImEx methods. The ImEx results use two splittings of \eqref{eq:ex-adv-diff-react}, in both cases diffusion is treated implicitly and advection explicitly while the reactions are treated implicitly (ImEx-1) or explicitly (ImEx-2). Each method is paired with the trivial (T), max order (M), variable order (V), and cutoff (C) predictors. The statistics given are the number of successful steps (Steps), number of failed steps due to an error test or nonlinear solver failure (Err and Solve fails), number of explicit and implicit function evaluations ($f^E$ and $f^I$ evals), number of nonlinear solver iterations and failures (NLS iters and fails), number of linear solver setups (LS setups), and number of Jacobian evaluations ($J$ evals). \editmade{Additionally, the maximum relative error at the final time compared to a reference solution is also given.} \label{t:ark-results} }
\begin{tabular}{ lrrrr|rrrr|rrrr }
\toprule
& \multicolumn{4}{c|}{DIRK} & \multicolumn{4}{c|}{ImEx-1} & \multicolumn{4}{c}{ImEx-2} \\
& \multicolumn{4}{c|}{Predictor} & \multicolumn{4}{c|}{Predictor} &\multicolumn{4}{c}{Predictor} \\
\multicolumn{1}{c}{Stats} &
\multicolumn{1}{c}{T} & \multicolumn{1}{c}{M} & \multicolumn{1}{c}{V} & \multicolumn{1}{c|}{C} & \multicolumn{1}{c}{T} & \multicolumn{1}{c}{M} & \multicolumn{1}{c}{V} & \multicolumn{1}{c|}{C} & \multicolumn{1}{c}{T} & \multicolumn{1}{c}{M} & \multicolumn{1}{c}{V} & \multicolumn{1}{c}{C} \\
\midrule
Steps       &   34 &    21 &   24 &   24 &   31 &   21 &   25 &   25 &   255 &   258 &   256 &   250  \\
Err fails   &    0 &     0 &    0 &    0 &    0 &    0 &    0 &    0 &    56 &    67 &    62 &    44  \\
Solve fails &    6 &     0 &    1 &    1 &    4 &    0 &    1 &    1 &     0 &     0 &     0 &     0  \\
$f^E$ evals &    - &     - &    - &    - &  204 &  129 &  158 &  158 & 1,869 & 1,953 & 1,911 & 1,767  \\
$f^I$ evals &  758 &   385 &  460 &  487 &  672 &  385 &  475 &  495 & 3,424 & 3,578 & 3,501 & 3,237  \\
NLS iters   &  528 &   256 &  308 &  335 &  468 &  256 &  317 &  337 & 1,555 & 1,625 & 1,590 & 1,470  \\
NLS fails   &   24 &     6 &   10 &   14 &   18 &    6 &   10 &   13 &     0 &     0 &     0 &     0  \\
LS setups   &   46 &    18 &   24 &   28 &   35 &   18 &   25 &   28 &   156 &   171 &   167 &   134  \\
$J$ evals   &   25 &     7 &   11 &   15 &   19 &    7 &   11 &   14 &    57 &    68 &    63 &    45  \\
\editmade{Error $\times10^{-4}$}  &
\editmade{$0.34$} &
\editmade{$1.7$} &
\editmade{$2.0$} &
\editmade{$1.8$} &
\editmade{$0.18$} &
\editmade{$2.5$} &
\editmade{$2.1$} &
\editmade{$2.2$} &
\editmade{$0.05$} &
\editmade{$7.2$} &
\editmade{$3.2$} &
\editmade{$0.1$} \\
\bottomrule
\end{tabular}
\end{table}





\editmade{Table \ref{t:ark-results} shows various integrator statistics for each configuration. In this test, where there are long periods of slow change followed by shorter, faster transition intervals, predictors using higher order extrapolation can increase efficiency compared to the trivial predictor. In the DIRK and IMEX-1 cases, the max order predictor provides significant reductions (32\% or more) in all integrator statistics while producing results close to the requested tolerance. The variable order and cutoff predictors also improve performance in the DIRK and IMEX-1 cases but to a lesser degree (reductions of 19\% or more) and are better suited to cases where the max order predictor becomes unstable. For the ImEx-2 setup, where only a single Newton iteration is applied in the implicit solve and results are more sensitive to predictor error, the more conservative trivial and cutoff predictors have the best performance. The cutoff predictor gives the best results providing a small improvement (up to a 20\%) over the trivial predictor in all integrator statistics.}


Comparing the integration methods more broadly, the performance of the DIRK and ImEx-1 methods are nearly identical as they use the same implicit method.  We note that with ImEx-ARK methods, the terms included in $f^E$ are evaluated far less often than those in $f^I$, since $f^E$ is only evaluated once per stage rather than once per nonlinear iteration. As such, when $f^E$ is expensive but non-stiff, an ImEx approach can offer greater efficiency. Alternatively, when an efficient nonlinear solver is unavailable for the unsplit right-hand side, an ImEx approach may offer similar performance without requiring an algebraic solver for the full system. Considering ImEx-2, we see that it is far more expensive than the DIRK or ImEx-1 approaches since its time steps are determined by the reactions.  However in terms of $f^E$ evaluations, ImEx-2 is on par with the advection-reaction results using ERKStep, as its implicit treatment of diffusion successfully alleviates the step size restrictions that would be present if this problem were treated fully explicitly.


\subsubsection{MRIStep}
\label{sec:mristep_example}

With ARKStep, \eqref{eq:ex-adv-diff-react} was split into implicit and explicit partitions using the same step size for all components, but, as was clear from Table \ref{t:ark-results}, resolving the reactions required smaller step sizes than advection and diffusion.  Therefore, we apply the default third-order ImEx-MRI-GARK method (IMEX-MRI-GARK3a from \cite{CR2021}) in MRIStep with advection and diffusion at the slow time scale using a fixed step size of $H=0.1$, and the reactions at the fast time scale. For this fast scale we consider three integration methods: the default third-order adaptive step size ERK method \cite{BS1989} in ARKStep that will resolve the \editmade{stiffest reaction time scale}, the default third-order adaptive step size DIRK method (ARK3(2)4L[2]SA–ESDIRK from \cite{KC2003}) in ARKStep that will step over the \editmade{stiffest reaction time scale} but at an increased cost per step, or an adaptive order and step size BDF method from CVODE that will also step over the \editmade{stiffest reaction time scale but} with a lower cost per step.  At the slow time scale, the linearly implicit systems are solved with a single Newton iteration and a banded direct linear solver. When using the DIRK and BDF methods at the fast time scale, the nonlinear implicit systems are solved with a modified Newton iteration paired with a banded direct linear solver. The integrators all use the same relative and absolute tolerances of $10^{-4}$ and $10^{-9}$, respectively.

To use BDF methods for the fast integration, the example code wraps CVODE as an MRIStepInnerStepper object.  The derived class implemented in the example code contains the data needed by the fast integrator (e.g., the CVODE memory structure) and defines three methods utilized by MRIStep: \texttt{Evolve}, \texttt{FullRHS}, and \texttt{Reset}. The first evolves the fast IVP system from step \ref{step:fastIVP} of the MRI method over a given time interval, the second evaluates $f^f(t,y)$ for slow time scale dense output, and the third resets the integrator to a given time and state value.  \editmade{We note that the linear multistep methods in CVODE bootstrap up from first order following each \texttt{Reset} and, unlike the Runge-Kutta methods, do not retain the error history across fast integrations. To prevent overly aggressive step size changes in these bootstrap phases, we reduce the maximum step size change after the first internal step from $10^4$ to $10$, the default value for subsequent steps.}

\begin{table}[htb!]
\centering
\caption{\label{t:mri-results} Integrator statistics using the default 3rd order ImEx MRI method paired with either an ERK method, DIRK method, or BDF method for the fast time scale integrator. The statistics given are the number of slow time steps (Steps slow), number of successful fast time steps (Steps fast), number of failed fast time steps (Failed steps fast), number of slow explicit, slow implicit, and fast function evaluations ($f^E$, $f^I$, and $f^f$ evals), number of nonlinear iterations and failures at the slow and fast time scales (NLS iters and fails), number of slow and fast linear solver setups (LS setups), and number of slow and fast Jacobian evaluations ($J$ evals). \editmade{Additionally, the maximum relative error at the final time compared to a reference solution is also given.}}
\begin{tabular}{ lrrr }
\toprule
\multicolumn{1}{c}{Stats} & \multicolumn{1}{c}{MRI-ERK} & \multicolumn{1}{c}{MRI-DIRK} & \multicolumn{1}{c}{MRI-BDF} \\
\midrule
Steps slow        &   100 &   100 &   100 \\
Steps fast        &   426 &   307 & \editmade{1,709} \\
Failed steps fast &     1 &     0 & \editmade{0} \\
$f^E$ evals       &   401 &   401 &   401 \\
$f^I$ evals       &   701 &   701 &   701 \\
$f^f$ evals       & 2,111 & 3,470 & \editmade{2,620} \\
Slow NLS iters    &   300 &   300 &   300 \\
Slow NLS fails    &     0 &     0 &     0 \\
Slow LS setups    &     1 &     1 &     1 \\
Slow $J$ evals    &     1 &     1 &     1 \\
Fast NLS iters    &    -- & 1,839 & \editmade{1,720} \\
Fast NLS fails    &    -- &     0 &     0 \\
Fast LS setups    &    -- &   304 & \editmade{1,636} \\
Fast $J$ evals    &    -- &   300 &   401 \\
\editmade{Error $\times10^{-4}$} & \editmade{$0.44$} & \editmade{$0.01$} & \editmade{$1.3$} \\
\bottomrule
\end{tabular}
\end{table}

Table \ref{t:mri-results} shows various integrator statistics for each configuration. Since all three setups apply the same ImEx method at the slow time scale, the ``slow'' statistics are identical. Comparing the performance statistics from the fast integration we see various trade-offs between the methods. The DIRK method \editmade{has the lowest error, significantly over-solving the problem, and} takes the fewest steps (approximately one per fast solve), but it requires \editmade{the most} fast function evaluations due to the \editmade{multiple} nonlinear solves in each step. The ERK method takes more steps \editmade{and has a larger error but} requires fewer function evaluations, making it the most efficient for this reaction model. \editmade{Finally, the BDF method gives a result close to the requested tolerance and takes the most steps (approximately six per fast solve) as it must bootstrap up from first order. However, the method has 24\% fewer fast function evaluations than the DIRK method since only one nonlinear solve is required per step. Thus, for stiffer reaction mechanisms with a greater separation between the fast and slow time scales, the BDF method may become the best approach.}

For any given application, the optimal choice of integrator family (explicit, ImEx, multirate), problem splittings (implicit-explicit and/or fast-slow), Runge--Kutta or MRI method coefficients, algebraic solvers, adaptivity methods, and solver parameters will obviously be problem specific. However, this sequence of examples illustrates that through ARKODE's broad range of options and simple API, users may easily explore these choices to discover what works best for their application.

\subsection{\editmade{Large-Scale 3D Demonstration Problem}}
\label{sec:demo}

As a final demonstration of ARKODE's flexibility and performance, we consider the three-dimensional nonlinear inviscid compressible Euler equations, with advection and reaction of chemical species,
\begin{equation}
    \label{eq:compressibleEuler}
    w_t = -\nabla\cdot F(w) + R(w),
\end{equation}
with the independent variables $(X,t) = (x,y,z,t) \in \Omega \times [0, t_f]$, and where the spatial domain is a three-dimensional cube, $\Omega = [0, 1] \times [0, 1] \times [0, 1]$. The solution is given by $w = [ \rho\; \rho v_x\; \rho v_y\; \rho v_z\; e_t\; \mathbf{c} ]^T$, corresponding to the density, momentum in the x, y, and z directions, total energy per unit volume, and some number of chemical densities $\mathbf{c}\in\mathbb{R}^{nchem}$ that are advected along with the fluid. The fluxes are given by
\begin{align}
    \label{eq:EulerFluxes}
    F_x(w) &= \begin{bmatrix} \rho v_x & \rho v_x^2 + p & \rho v_x v_y & \rho v_x v_z & v_x (e_t+p) & \mathbf{c} v_x \end{bmatrix}^T,\\
    F_y(w) &= \begin{bmatrix} \rho v_y & \rho v_x v_y & \rho v_y^2 + p & \rho v_y v_z & v_y (e_t+p) & \mathbf{c} v_y \end{bmatrix}^T,\\
    F_z(w) &= \begin{bmatrix} \rho v_z & \rho v_x v_z & \rho v_y v_z & \rho v_z^2 + p & v_z (e_t+p) & \mathbf{c} v_z \end{bmatrix}^T.
\end{align}
The reaction \editmade{mechanism} is encoded in $R(w)$, and the ideal gas equation of state gives $p = (\gamma-1) (e_t - \frac{\rho}{2} (v_x^2 + v_y^2 + v_z^2))$, where $\gamma = c_p/c_v$ is the \editmade{constant} ratio of specific heats for the gas.

We discretize this problem using the method of lines, where we first semi-discretize in space using a regular finite volume grid with dimensions $n_x \times n_y \times n_z$, with fluxes at cell faces calculated using a 5th-order FD-WENO reconstruction \cite{Shu2003}.  MPI parallelization is achieved using a standard 3D domain decomposition of $\Omega$.  We organize the corresponding spatially-discretized solution vector, $y$, using the SUNDIALS MPIManyVector \cite{gardner2020enabling} that wraps node-local vectors for each of the fields, $\rho, \ldots, \mathbf{c}$, to create the overall state vector, $w$, and provides the requisite MPI functionality for coordinating vector operations among the subvectors. The hydrodynamic fields, ($\rho$, $\rho\mathbf{v}$, and $e$), are stored in CPU memory, and the chemical densities, $\mathbf{c}$, are stored in GPU memory.  After spatial semi-discretization, we are faced with a large IVP system,
\begin{equation}
    \label{eq:demonstration_IVP}
    w'(t) = f_1(w) + f_2(w), \quad w(t_0)=w_0,
\end{equation}
where $f_1(w)$ and $f_2(w)$ contain the spatially discretized forms of $-\nabla\cdot F(w)$ and $R(w)$, respectively.  For additional information on this test problem or our computational implementation, please see the technical report \cite{reynoldsSUNDIALSMultiphysicsMPIManyVector2019a} and the GitHub repository \cite[v4.0]{RGB2021}.

In the results that follow, we simulate the advection and reaction of a low density primordial gas, present in models of the early universe \cite{Abel1997,Glover2008,Kreckel2010,Trevisan_2002}.  This gas consists of 10 advected species (i.e., $nchem=10$), corresponding to various ionization states of atomic and molecular Hydrogen and Helium, free electrons, and the internal gas energy.  For this setup the reaction \editmade{mechanism} is considerably stiffer than the advection, so we consider two forms for temporal evolution:
\begin{itemize}
\item[(a)] $\bigO{H^4}$, temporally-adaptive, single-step, ImEx method from ARKStep (ARK4(3)7L[2]SA$_1$ from \cite{KC2019}), where $f_1$ is treated explicitly and $f_2$ is treated implicitly, and
\item[(b)\label{i:demo-mri}] $\bigO{H^3}$ explicit MIS method from MRIStep (using the ``RK3'' coefficients from \cite{KW1998}), where $f_1$ is treated at the slow time scale using fixed step sizes, and $f_2$ is treated at the fast time scale using the DIRK portion of the $\bigO{H^4}$, adaptive, DIRK method from (a).
\end{itemize}
For (a) we use SUNDIALS' modified Newton solver to handle the global nonlinear algebraic systems arising at each implicit stage of each time step.  Since only $f_2$ is treated implicitly and the reactions are purely local in space, the Newton linear systems are block-diagonal. As such, we provide a custom \texttt{SUNLinearSolver} implementation that solves each MPI rank-local linear system independently. The portion of the Jacobian matrix on each rank, $J_p$, is itself block-diagonal, i.e.,
\begin{equation*}
  J_p = \begin{bmatrix} J_{p,1,1,1} & & & \\
  & J_{p,2,1,1} & & \\
  & & \ddots & \\
  & & & J_{p,n_{xloc},n_{yloc},n_{zloc}} \end{bmatrix},
\end{equation*}
with each cell-local block $J_{p,i,j,k} \in \R^{10\times 10}$. We further leverage this structure by solving each $J_p x_p = b_p$ linear system using the SUNDIALS \texttt{SUNLinearSolver} implementation interfacing to GPU-enabled, batched, dense LU solver routines from MAGMA \cite{tdb10}.

The multirate approach (b) can leverage the structure of $f_2$ at a higher level.  Since the MRI method applied to this problem evolves ``fast'' sub-problems of the form
\begin{equation}
\label{eq:demonstration_fastlocal}
   v'(t) = f_2(t,v) + r_i(t), \quad i=2,\ldots,s,
\end{equation}
and all MPI communication necessary to construct the forcing functions, $r_i(t)$, has already been performed, each sub-problem \eqref{eq:demonstration_fastlocal} consists of $n_x\,n_y\,n_z$ spatially-decoupled fast IVPs. We construct a custom fast integrator that groups all the independent fast IVPs on an MPI rank together as a single system evolved using a rank-local ARKStep instance.  The code for this custom integrator itself is minimal, primarily consisting of steps to access the local subvectors in $w$ on a given MPI rank and wrapping them in MPI-unaware ManyVectors provided to the local ARKStep instance. The collection of independent local IVPs also leads to a block diagonal Jacobian, and we again utilize the SUNDIALS interface to MAGMA for solving linear systems within the modified Newton iteration.

These tests use periodic boundary conditions and a ``clumpy'' initial background density field
\begin{equation*}
    \rho_b(X) = 100 \left(1 + \sum_{i=1}^{N_c} s_i \exp\left(-2 \left(\frac{\|X-X_i\|}{r_i}\right)^2\right)\right),
\end{equation*}
where $s_i \in [0,5]$, $r_i \in [3\Delta x,6\Delta x]$, and $X_i\in \Omega_p$ are uniformly-distributed random values and there are 10 ``clumps'' per MPI rank. The initial background temperature is $T_b(X)= 10$, and the initial velocity field is identically zero. The initial conditions for density and temperature are obtained by adding a Gaussian bump to the background states
\begin{equation*}
    \rho_0(X) = \rho_b(X) + 500 \exp\left(-2 \left(\frac{\|X-0.5\|}{0.1}\right)^2 \right),
    \quad
    T_0(X) = T_b(X) + 50 \exp\left(-2\left(\frac{\|X-0.5\|}{0.1}\right)^2 \right),
\end{equation*}
causing a high-pressure front that emanates from the center of the domain.  We initialize the chemical \editmade{species} to consist almost exclusively of neutral Hydrogen (76\%) and Helium (24\%), with trace amounts of ionized Hydrogen and Helium (0.0000000001\%) outside of this high pressure region.  As the region expands and increases the surrounding temperature, the chemical concentrations in those cells rapidly transition to a different equilibrium, resulting in a chemical time scale that is multiple orders of magnitude faster than the spatial propagation of the high-pressure front.

\begin{figure}[htb!]
    \centering
    \begin{subfigure}[b]{0.45\textwidth}
        \centering
        \includegraphics[width=\textwidth,trim={1.0cm 0.0cm 2.5cm 0.75cm},clip]{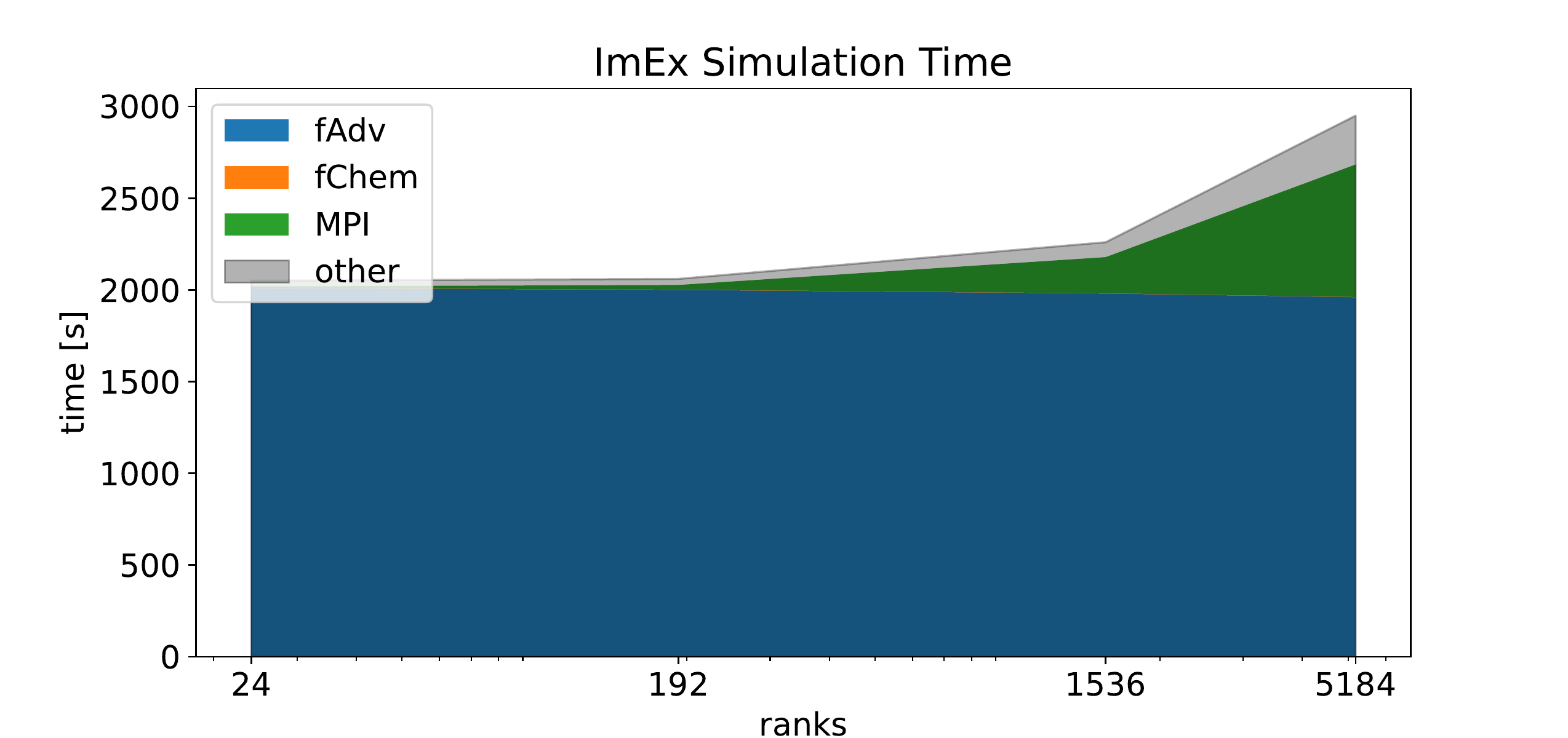}
        \label{fig:imex_simtimes}
    \end{subfigure}
    \hfill
    \begin{subfigure}[b]{0.45\textwidth}
       \centering
        \includegraphics[width=\textwidth,trim={1.0cm 0.0cm 2.5cm 0.75cm},clip]{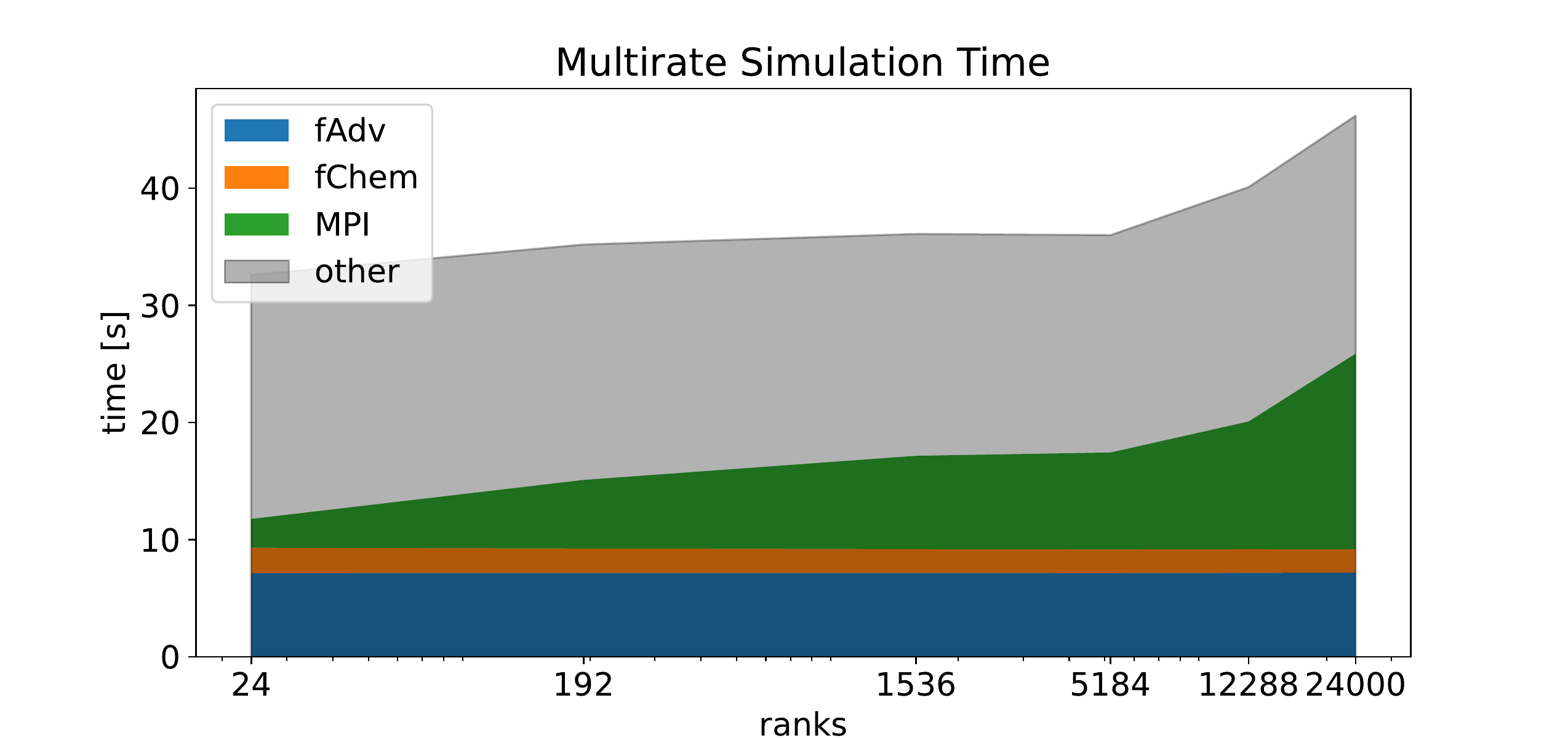}
        \label{fig:mr_simtimes}
    \end{subfigure}

    \begin{subfigure}[b]{0.45\textwidth}
        \centering
        \includegraphics[width=\textwidth,trim={1.0cm 0.0cm 2.5cm 0.75cm},clip]{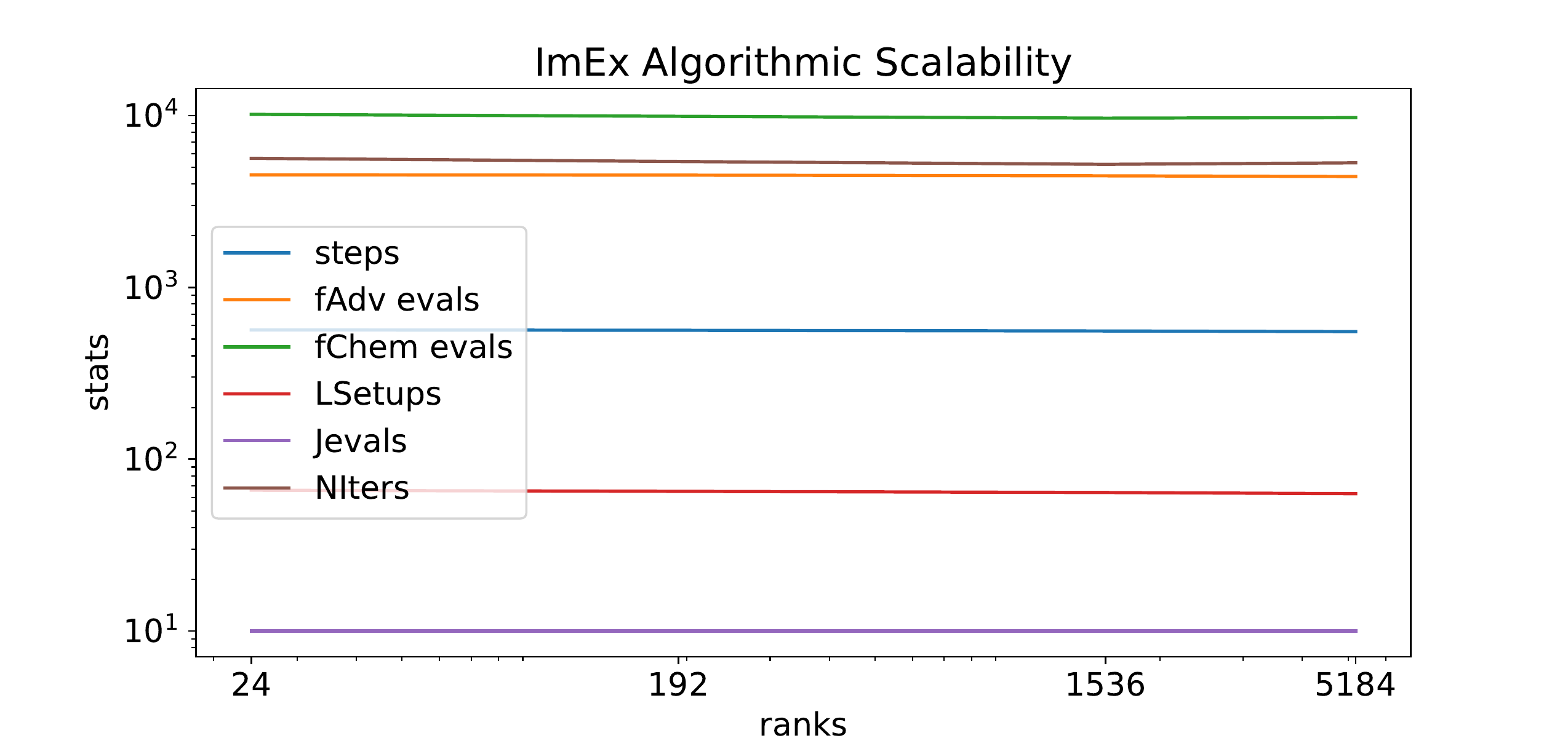}
        \label{fig:imex_stats}
    \end{subfigure}
    \hfill
    \begin{subfigure}[b]{0.45\textwidth}
       \centering
        \includegraphics[width=\textwidth,trim={1.0cm 0.0cm 2.5cm 0.75cm},clip]{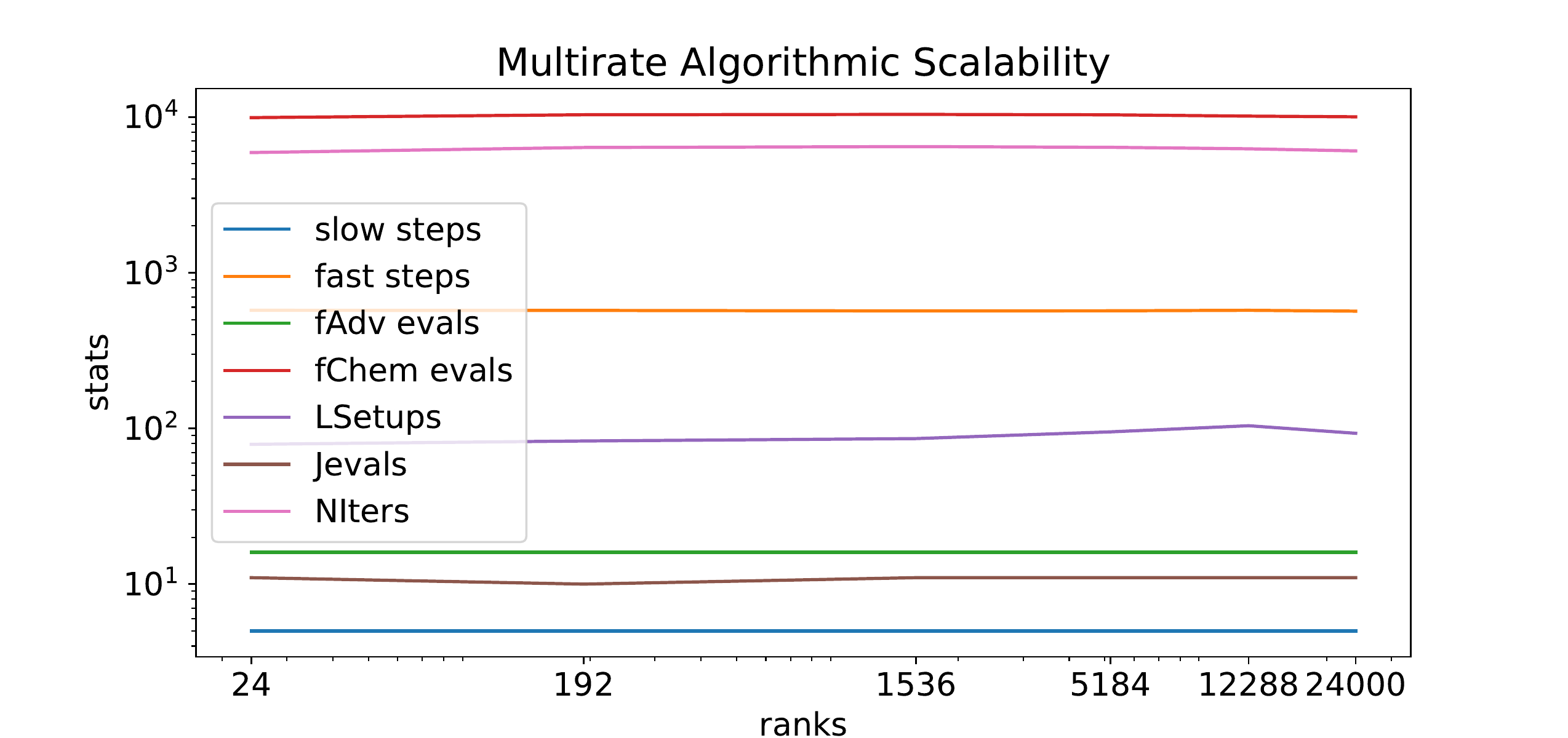}
        \label{fig:mr_stats}
    \end{subfigure}
    \caption{Demonstration code scaling: simulation times (first row) and algorithmic scalability (second row).  \texttt{fAdv} corresponds to evaluation of the advection operator and \texttt{fChem} the chemistry operator.  In the first row, \texttt{MPI} corresponds with the global synchronization times required following each rank-local solver component, and \texttt{other} contains all components that are not directly measured (e.g., time spent within the GPU-based MAGMA linear solver).  In the second row, \texttt{steps}, \texttt{slow steps} and \texttt{fast steps} show the total number of time steps taken at each level of each solver, \texttt{LSetups} shows the total number of linear solver setup calls, \texttt{Jevals} shows the total number of Jacobian evaluations, and \texttt{Niters} shows the total number of Newton iterations.}
    \label{fig:demonstration_results}
\end{figure}

In Figure \ref{fig:demonstration_results} we present timing results for this problem run on Summit at the Oak Ridge Leadership Computing Facility. \editmade{Each Summit node contains two IBM POWER9 processors and six NVIDIA Tesla V100 GPUs. The nodes are connected to a dual-rail EDR InfiniBand network in a non-block fat tree topology.} We performed a weak-scaling simulation with:
\begin{itemize}
    \item a $25 \times 25 \times 25$ spatial grid per MPI rank,
    \item multirate integrators using a slow time step size satisfying the CFL constraint $H \le 10\Delta x$,
    \item fast multirate time step sizes and ImEx time step sizes chosen adaptively by the ARKStep integrator with relative and absolute tolerances of $10^{-5}$ and $10^{-9}$, respectively,
    \item temporal evolution over $[0,5H]$, and
    \item 6 MPI ranks per Summit node, with one NVIDIA Volta GPU tied to each MPI rank.
\end{itemize}
\editmade{We note that both the ImEx and MRI approaches select time steps to accurately track the chemical evolution.  In the ImEx case, the entire simulation evolves with a shared step that is therefore set by the chemistry in the ``hardest'' spatial cell.  However, in the multirate case, only the fast time scale must track the chemistry, and that occurs on a rank-by-rank basis, allowing other MPI ranks and other physical processes to evolve with larger time steps.}

The top row of Figure \ref{fig:demonstration_results} shows the total simulation times for both configurations of the problem, while the second row shows the corresponding algorithmic scalability.  Examining the second row, we see that for all solver statistics both configurations demonstrate ideal algorithmic scalability, implying that our algorithm choices map optimally to the underlying problem structure.

From the first row of Figure \ref{fig:demonstration_results} two observations are immediately clear.  First, the high cost associated with each evaluation of the advection operator (shown as \texttt{fAdv}) causes the MRI approach to easily outperform the ImEx method, due to its much smaller number of evaluations.  Second, the primary runtime slowdown experienced within both approaches arises in the green \texttt{MPI} component.  For the ImEx method, this component corresponds solely to a single \texttt{MPI\_Allreduce} after each rank-local linear solve to determine the overall ``success'' or ``failure'' of the global linear solve.  Thus, the MPI cost reflects the cost of global synchronization.  Similarly, in the MRI case the \texttt{MPI} component measures an \texttt{MPI\_Allreduce} that determines success of the fast local IVP solves.  Hence, the fundamental scalability difference between the ImEx and MRI approaches results from the frequency of these synchronizations. For the ImEx method synchronizations occur following each linear solve, within each Newton iteration, within each implicit stage, within each time step.  Since ARKStep's temporal adaptivity tracked the (fast) reaction time scale, the simulation required a \emph{large} number of synchronizations.  On the other hand, with the MRI method these synchronizations occur at the slow time scale, i.e., considerably less often than in the ImEx case.

Taking a step back to focus on ARKODE as an infrastructure, we conclude that its algorithmic flexibility, and in particular its support for custom solver components and the new MRIStep multirate module, enables a rich exploration of the ``solver space'' for this and many problems.

\section{Conclusions}
\label{sec:conclusions}

ARKODE is \editmade{a} highly flexible, usable, and scalable infrastructure for one-step IVP solvers.  It is currently used by a variety of applications with functionality that is rapidly expanding to encompass novel and flexible numerical methods for simulating complex phenomena.  The current version of ARKODE is \editmade{5.5.0}, is contained within SUNDIALS version \editmade{6.5.0}, and is widely available through distribution modes including Spack \cite{gamblin+:sc15} and \href{https://github.com/LLNL/sundials}{GitHub}.

We have designed ARKODE to support a range of one-step methods, providing reusable infrastructure components that new methods may leverage to minimize the time between initial inception and public release in a high-quality open-source software library. By supplying a method-specific \emph{step} function that propagates the IVP solution and provides an estimate of the local truncation error, ARKODE can wrap these methods to provide well-established usage modes that evolve an IVP over a time interval or return after each internal step.  ARKODE also provides added features including high-order dense output, event handling, and support for inequality constraints on solution components.

The existing ecosystem of ARKODE time stepping modules allows users to explore various options for time integration with only minor modifications to their codes, allowing these algorithms to increase in complexity with their application.  The current modules include a lean explicit Runge--Kutta time stepper, an additive Runge--Kutta time stepper for explicit, implicit, and ImEx integration, and a new multirate infinitesimal time stepper that supports explicit, implicit, and ImEx integration of at slow time scale while allowing for arbitrary integration methods at the fast time scale.  \editmade{Through its support for user-supplied components (e.g., Butcher tables, MRI coefficients, and algebraic solver modules) within its existing time-stepping modules, as well as its infrastructure support for generic time-stepping modules, ARKODE can serve as a robust platform for rapidly developing and testing ``experimental'' methods at all levels of the solver hierarchy.}

As with most open-source packages, we continue to add functionality to ARKODE.  Current enhancements include support for multirate temporal adaptivity within MRIStep, as well as new time stepping modules for advanced one-step methods (positivity/structure preserving, etc.).

\begin{acks}
The authors would like to thank Alan Hindmarsh, Cody Balos, Jean Sexton, Slaven Peles, Ting Yan, and John Loffeld for their software and mathematical contributions.  We would also like to thank Gregg Hommes, Sylvie Aubry, Tom Arsenlis, Paul Ullrich, Jorge Guerra, Chris Vogl, Andrew Steyer, Mark Taylor, Darin Ernst, and Manaure Francisquez for their use and feedback on ARKODE, and their patience as we ironed out kinks and constructed improvements.

Support for this work was provided in part by the \grantsponsor{SciDAC}{U.S. Department of Energy, Office of Science, Office of Advanced Scientific Computing Research, Scientific Discovery through Advanced Computing (SciDAC) Program}{https://www.scidac.gov/} through the Frameworks, Algorithms, and Scalable Technologies for Mathematics (FASTMath) Institute, under Lawrence Livermore National Laboratory subcontract \grantnum{SciDAC}{B626484} and DOE award \grantnum{SciDAC}{DE-SC0021354}.

This research was supported in part by the \grantsponsor{ECP}{Exascale Computing Project}{https://exascaleproject.org/} (\grantnum{ECP}{17-SC-20-SC}), a collaborative effort of the U.S. Department of Energy Office of Science and the National Nuclear Security Administration.

This research used resources of the Oak Ridge Leadership Computing Facility at the Oak Ridge National Laboratory, which is supported by the Office of Science of the U.S. Department of Energy under Contract No. DE-AC05-00OR22725.

This work was performed under the auspices of the U.S. Department of Energy by Lawrence Livermore National Laboratory under contract DE-AC52-07NA27344. Lawrence Livermore National Security, LLC. LLNL-JRNL-835429.
\end{acks}

\bibliographystyle{ACM-Reference-Format}
\bibliography{sources}


\begin{thebibliography}{66}


\ifx \showCODEN    \undefined \def \showCODEN     #1{\unskip}     \fi
\ifx \showDOI      \undefined \def \showDOI       #1{#1}\fi
\ifx \showISBNx    \undefined \def \showISBNx     #1{\unskip}     \fi
\ifx \showISBNxiii \undefined \def \showISBNxiii  #1{\unskip}     \fi
\ifx \showISSN     \undefined \def \showISSN      #1{\unskip}     \fi
\ifx \showLCCN     \undefined \def \showLCCN      #1{\unskip}     \fi
\ifx \shownote     \undefined \def \shownote      #1{#1}          \fi
\ifx \showarticletitle \undefined \def \showarticletitle #1{#1}   \fi
\ifx \showURL      \undefined \def \showURL       {\relax}        \fi
\providecommand\bibfield[2]{#2}
\providecommand\bibinfo[2]{#2}
\providecommand\natexlab[1]{#1}
\providecommand\showeprint[2][]{arXiv:#2}

\bibitem[Abel et~al\mbox{.}(1997)]%
        {Abel1997}
\bibfield{author}{\bibinfo{person}{Tom Abel}, \bibinfo{person}{Peter Anninos},
  \bibinfo{person}{Yu Zhang}, {and} \bibinfo{person}{Michael~L. Norman}.}
  \bibinfo{year}{1997}\natexlab{}.
\newblock \showarticletitle{Modeling primordial gas in numerical cosmology}.
\newblock \bibinfo{journal}{\emph{New Astronomy}} \bibinfo{volume}{2},
  \bibinfo{number}{3} (\bibinfo{year}{1997}), \bibinfo{pages}{181 -- 207}.
\newblock
\showISSN{1384-1076}
\urldef\tempurl%
\url{https://doi.org/10.1016/S1384-1076(97)00010-9}
\showDOI{\tempurl}


\bibitem[Abhyankar et~al\mbox{.}(2018)]%
        {ABCGSZ2018}
\bibfield{author}{\bibinfo{person}{Shrirang Abhyankar}, \bibinfo{person}{Jed
  Brown}, \bibinfo{person}{Emil~M. Constantinescu}, \bibinfo{person}{Debojyoti
  Ghosh}, \bibinfo{person}{Barry~F. Smith}, {and} \bibinfo{person}{Hong
  Zhang}.} \bibinfo{year}{2018}\natexlab{}.
\newblock \bibinfo{title}{PETSc/TS: A Modern Scalable ODE/DAE Solver Library}.
\newblock
\newblock
\showeprint[arxiv]{1806.01437}~[math.NA]


\bibitem[Anderson et~al\mbox{.}(2021)]%
        {mfem}
\bibfield{author}{\bibinfo{person}{Robert Anderson}, \bibinfo{person}{Julian
  Andrej}, \bibinfo{person}{Andrew Barker}, \bibinfo{person}{Jamie Bramwell},
  \bibinfo{person}{Jean-Sylvain Camier}, \bibinfo{person}{Jakub Cerveny},
  \bibinfo{person}{Veselin Dobrev}, \bibinfo{person}{Yohann Dudouit},
  \bibinfo{person}{Aaron Fisher}, \bibinfo{person}{Tzanio Kolev},
  \bibinfo{person}{Will Pazner}, \bibinfo{person}{Mark Stowell},
  \bibinfo{person}{Vladimir Tomov}, \bibinfo{person}{Ido Akkerman},
  \bibinfo{person}{Johann Dahm}, \bibinfo{person}{David Medina}, {and}
  \bibinfo{person}{Stefano Zampini}.} \bibinfo{year}{2021}\natexlab{}.
\newblock \showarticletitle{{MFEM}: A Modular Finite Element Methods Library}.
\newblock \bibinfo{journal}{\emph{Computers \& Mathematics with Applications}}
  \bibinfo{volume}{81} (\bibinfo{year}{2021}), \bibinfo{pages}{42--74}.
\newblock
\urldef\tempurl%
\url{https://doi.org/10.1016/j.camwa.2020.06.009}
\showDOI{\tempurl}


\bibitem[Balos et~al\mbox{.}(2021)]%
        {balosEnablingGPUAccelerated2021}
\bibfield{author}{\bibinfo{person}{Cody~J. Balos}, \bibinfo{person}{David~J.
  Gardner}, \bibinfo{person}{Carol~S. Woodward}, {and}
  \bibinfo{person}{Daniel~R. Reynolds}.} \bibinfo{year}{2021}\natexlab{}.
\newblock \showarticletitle{Enabling {{GPU}} Accelerated Computing in the
  {{SUNDIALS}} Time Integration Library}.
\newblock \bibinfo{journal}{\emph{Parallel Comput.}}  \bibinfo{volume}{108}
  (\bibinfo{date}{Dec.} \bibinfo{year}{2021}), \bibinfo{pages}{102836}.
\newblock
\showISSN{01678191}
\urldef\tempurl%
\url{https://doi.org/10.1016/j.parco.2021.102836}
\showDOI{\tempurl}


\bibitem[Bogacki and Shampine(1989)]%
        {BS1989}
\bibfield{author}{\bibinfo{person}{Przemyslaw Bogacki} {and}
  \bibinfo{person}{Lawrence~F. Shampine}.} \bibinfo{year}{1989}\natexlab{}.
\newblock \showarticletitle{A 3(2) pair of Runge–Kutta formulas}.
\newblock \bibinfo{journal}{\emph{Appl.~Math.~Lett.}}  \bibinfo{volume}{2}
  (\bibinfo{year}{1989}), \bibinfo{pages}{321--325}.
\newblock
\urldef\tempurl%
\url{https://doi.org/10.1016/0893-9659(89)90079-7}
\showDOI{\tempurl}


\bibitem[Cash and Karp(1990)]%
        {CK1990}
\bibfield{author}{\bibinfo{person}{Jeff~R. Cash} {and} \bibinfo{person}{Alan~H.
  Karp}.} \bibinfo{year}{1990}\natexlab{}.
\newblock \showarticletitle{A variable order Runge-Kutta method for initial
  value problems with rapidly varying right-hand sides}.
\newblock \bibinfo{journal}{\emph{{ACM} Trans. Math. Software}}
  \bibinfo{volume}{16} (\bibinfo{year}{1990}), \bibinfo{pages}{201--222}.
\newblock
\urldef\tempurl%
\url{https://doi.org/10.1145/79505.79507}
\showDOI{\tempurl}


\bibitem[Cervi and Spiteri(2019)]%
        {cervi2019comparison}
\bibfield{author}{\bibinfo{person}{Jessica Cervi} {and}
  \bibinfo{person}{Raymond~J Spiteri}.} \bibinfo{year}{2019}\natexlab{}.
\newblock \showarticletitle{A comparison of fourth-order operator splitting
  methods for cardiac simulations}.
\newblock \bibinfo{journal}{\emph{Applied Numerical Mathematics}}
  \bibinfo{volume}{145} (\bibinfo{year}{2019}), \bibinfo{pages}{227--235}.
\newblock


\bibitem[Chinomona and Reynolds(2021)]%
        {CR2021}
\bibfield{author}{\bibinfo{person}{Rujeko Chinomona} {and}
  \bibinfo{person}{Daniel~R. Reynolds}.} \bibinfo{year}{2021}\natexlab{}.
\newblock \showarticletitle{Implicit-{Explicit} {Multirate} {Infinitesimal}
  {GARK} {Methods}}.
\newblock \bibinfo{journal}{\emph{SIAM J. Sci. Comput.}} \bibinfo{volume}{43},
  \bibinfo{number}{5} (\bibinfo{date}{Jan.} \bibinfo{year}{2021}),
  \bibinfo{pages}{A3082--A3113}.
\newblock
\urldef\tempurl%
\url{https://doi.org/10.1137/20M1354349}
\showDOI{\tempurl}


\bibitem[Constantinescu and Sandu(2013)]%
        {constantinescuExtrapolatedMultirateMethods2013}
\bibfield{author}{\bibinfo{person}{Emil~M. Constantinescu} {and}
  \bibinfo{person}{Adrian Sandu}.} \bibinfo{year}{2013}\natexlab{}.
\newblock \showarticletitle{Extrapolated {{Multirate Methods}} for
  {{Differential Equations}} with {{Multiple Time Scales}}}.
\newblock \bibinfo{journal}{\emph{J Sci Comput}} \bibinfo{volume}{56},
  \bibinfo{number}{1} (\bibinfo{date}{July} \bibinfo{year}{2013}),
  \bibinfo{pages}{28--44}.
\newblock
\showISSN{0885-7474, 1573-7691}
\urldef\tempurl%
\url{https://doi.org/10.1007/s10915-012-9662-z}
\showDOI{\tempurl}


\bibitem[Cooper and Sayfy(1980)]%
        {cooperAdditiveMethodsNumerical1980}
\bibfield{author}{\bibinfo{person}{Graeme~J. Cooper} {and} \bibinfo{person}{Ali
  Sayfy}.} \bibinfo{year}{1980}\natexlab{}.
\newblock \showarticletitle{Additive {{Methods}} for the {{Numerical Solution}}
  of {{Ordinary Differential Equations}}}.
\newblock \bibinfo{journal}{\emph{Math. Comp.}}  \bibinfo{volume}{35}
  (\bibinfo{year}{1980}), \bibinfo{pages}{1159--1172}.
\newblock


\bibitem[DeWitt et~al\mbox{.}(2022)]%
        {doecode_69332}
\bibfield{author}{\bibinfo{person}{Stephen DeWitt}, \bibinfo{person}{Philip
  Fackler}, \bibinfo{person}{Younggil Song}, \bibinfo{person}{Balasubramaniam
  Radhakrishnan}, {and} \bibinfo{person}{Sarma Gorti}.}
  \bibinfo{year}{2022}\natexlab{}.
\newblock \bibinfo{title}{MEUMAPPS (C++ Version)}.
\newblock
\newblock
\urldef\tempurl%
\url{https://doi.org/10.11578/dc.20220114.1}
\showDOI{\tempurl}


\bibitem[et~al.(2022)]%
        {xbraid-package}
\bibfield{author}{\bibinfo{person}{Robert D.~Falgout et al.}}
  \bibinfo{year}{2022}\natexlab{}.
\newblock \bibinfo{title}{{XB}raid: Parallel multigrid in time}.
\newblock \bibinfo{howpublished}{\url{http://llnl.gov/casc/xbraid}}.
\newblock


\bibitem[Falgout et~al\mbox{.}(2014)]%
        {falgout2014parallel}
\bibfield{author}{\bibinfo{person}{Robert~D. Falgout},
  \bibinfo{person}{Stephanie Friedhoff}, \bibinfo{person}{Tzanio~V. Kolev},
  \bibinfo{person}{Scott~P. MacLachlan}, {and} \bibinfo{person}{Jacob~B.
  Schroder}.} \bibinfo{year}{2014}\natexlab{}.
\newblock \showarticletitle{Parallel time integration with multigrid}.
\newblock \bibinfo{journal}{\emph{SIAM Journal of Scientific Computing}}
  \bibinfo{volume}{36} (\bibinfo{year}{2014}), \bibinfo{pages}{C635--C661}.
\newblock


\bibitem[Fish and Reynolds(2022)]%
        {FishReynolds2022}
\bibfield{author}{\bibinfo{person}{Alex~C. Fish} {and}
  \bibinfo{person}{Daniel~R. Reynolds}.} \bibinfo{year}{2022}\natexlab{}.
\newblock \bibinfo{title}{Adaptive time step control for infinitesimal
  multirate methods}.
\newblock
\newblock
\urldef\tempurl%
\url{https://doi.org/10.48550/ARXIV.2202.10484}
\showDOI{\tempurl}


\bibitem[Francisquez et~al\mbox{.}(2021)]%
        {mushroom-aps-2021}
\bibfield{author}{\bibinfo{person}{Manaure Francisquez},
  \bibinfo{person}{Darin~R. Ernst}, \bibinfo{person}{Daniel~R. Reynolds}, {and}
  \bibinfo{person}{Cody~J. Balos}.} \bibinfo{year}{2021}\natexlab{}.
\newblock \showarticletitle{{63rd Annual Meeting of the APS Division of Plasma
  Physics - A 2D gyrofluid model for coupled toroidal ITG/ETG multiscale
  turbulence and its comparison to gyrokinetics}}. In
  \bibinfo{booktitle}{\emph{Bulletin of the {American} {Physical} {Society}}},
  Vol.~\bibinfo{volume}{66}. \bibinfo{publisher}{American Physical Society},
  \bibinfo{address}{Pittsburgh, PA}.
\newblock


\bibitem[Gamblin et~al\mbox{.}(2015)]%
        {gamblin+:sc15}
\bibfield{author}{\bibinfo{person}{Todd Gamblin}, \bibinfo{person}{Matthew~P.
  LeGendre}, \bibinfo{person}{Michael~R. Collette}, \bibinfo{person}{Gregory~L.
  Lee}, \bibinfo{person}{Adam Moody}, \bibinfo{person}{Bronis~R. de Supinski},
  {and} \bibinfo{person}{W.~Scott Futral}.} \bibinfo{year}{2015}\natexlab{}.
\newblock \showarticletitle{{The Spack Package Manager: Bringing order to HPC
  software chaos}}. In \bibinfo{booktitle}{\emph{Supercomputing 2015 (SC'15)}}.
  \bibinfo{publisher}{ACM/IEEE}, \bibinfo{address}{Austin, Texas}.
\newblock


\bibitem[Gardner et~al\mbox{.}(2018)]%
        {gardner2018implicit}
\bibfield{author}{\bibinfo{person}{David~J. Gardner}, \bibinfo{person}{Jorge~E.
  Guerra}, \bibinfo{person}{Fran{\c{c}}ois~P. Hamon},
  \bibinfo{person}{Daniel~R. Reynolds}, \bibinfo{person}{Paul~A. Ullrich},
  {and} \bibinfo{person}{Carol~S. Woodward}.} \bibinfo{year}{2018}\natexlab{}.
\newblock \showarticletitle{Implicit--explicit (IMEX) Runge--Kutta methods for
  non-hydrostatic atmospheric models}.
\newblock \bibinfo{journal}{\emph{Geosci. Model Dev.}} \bibinfo{volume}{11},
  \bibinfo{number}{4} (\bibinfo{year}{2018}), \bibinfo{pages}{1497--1515}.
\newblock


\bibitem[Gardner et~al\mbox{.}(2020)]%
        {gardner2020enabling}
\bibfield{author}{\bibinfo{person}{David~J. Gardner},
  \bibinfo{person}{Daniel~R. Reynolds}, \bibinfo{person}{Carol~S. Woodward},
  {and} \bibinfo{person}{Cody~J. Balos}.} \bibinfo{year}{2020}\natexlab{}.
\newblock \bibinfo{title}{Enabling New Flexibility in the SUNDIALS Suite of
  Nonlinear and Differential/Algebraic Equation Solvers}.
\newblock
\newblock
\showeprint[arxiv]{2011.10073}~[cs.MS]


\bibitem[Gardner et~al\mbox{.}(2015)]%
        {gardner2015implicit}
\bibfield{author}{\bibinfo{person}{David~J. Gardner}, \bibinfo{person}{Carol~S.
  Woodward}, \bibinfo{person}{Daniel~R. Reynolds}, \bibinfo{person}{Gregg
  Hommes}, \bibinfo{person}{Sylvie Aubry}, {and} \bibinfo{person}{A.~Tom
  Arsenlis}.} \bibinfo{year}{2015}\natexlab{}.
\newblock \showarticletitle{Implicit integration methods for dislocation
  dynamics}.
\newblock \bibinfo{journal}{\emph{Modelling and Simulation in Materials Science
  and Engineering}} \bibinfo{volume}{23}, \bibinfo{number}{2}
  (\bibinfo{year}{2015}), \bibinfo{pages}{025006}.
\newblock


\bibitem[Gear and Wells(1984)]%
        {gearMultirateLinearMultistep1984}
\bibfield{author}{\bibinfo{person}{C.~William Gear} {and}
  \bibinfo{person}{Daniel~R. Wells}.} \bibinfo{year}{1984}\natexlab{}.
\newblock \showarticletitle{Multirate Linear Multistep Methods}.
\newblock \bibinfo{journal}{\emph{BIT}} \bibinfo{volume}{24},
  \bibinfo{number}{4} (\bibinfo{date}{Dec.} \bibinfo{year}{1984}),
  \bibinfo{pages}{484--502}.
\newblock
\showISSN{1572-9125}
\urldef\tempurl%
\url{https://doi.org/10.1007/BF01934907}
\showDOI{\tempurl}


\bibitem[Glover and Abel(2008)]%
        {Glover2008}
\bibfield{author}{\bibinfo{person}{Simon C.~O. Glover} {and}
  \bibinfo{person}{Tom Abel}.} \bibinfo{year}{2008}\natexlab{}.
\newblock \showarticletitle{{Uncertainties in H2 and HD chemistry and cooling
  and their role in early structure formation}}.
\newblock \bibinfo{journal}{\emph{Monthly Notices of the Royal Astronomical
  Society}} \bibinfo{volume}{388}, \bibinfo{number}{4} (\bibinfo{date}{08}
  \bibinfo{year}{2008}), \bibinfo{pages}{1627--1651}.
\newblock
\showISSN{0035-8711}
\urldef\tempurl%
\url{https://doi.org/10.1111/j.1365-2966.2008.13224.x}
\showDOI{\tempurl}


\bibitem[G{\"u}nther and Sandu(2016)]%
        {guntherMultirateGeneralizedAdditive2016}
\bibfield{author}{\bibinfo{person}{Michael G{\"u}nther} {and}
  \bibinfo{person}{Adrian Sandu}.} \bibinfo{year}{2016}\natexlab{}.
\newblock \showarticletitle{Multirate Generalized Additive {{Runge Kutta}}
  Methods}.
\newblock \bibinfo{journal}{\emph{Numer. Math.}} \bibinfo{volume}{133},
  \bibinfo{number}{3} (\bibinfo{date}{July} \bibinfo{year}{2016}),
  \bibinfo{pages}{497--524}.
\newblock
\showISSN{0029-599X, 0945-3245}
\urldef\tempurl%
\url{https://doi.org/10.1007/s00211-015-0756-z}
\showDOI{\tempurl}


\bibitem[Gustafsson(1991)]%
        {G1991}
\bibfield{author}{\bibinfo{person}{Kjell Gustafsson}.}
  \bibinfo{year}{1991}\natexlab{}.
\newblock \showarticletitle{Control theoretic techniques for stepsize selection
  in explicit Runge-Kutta methods}.
\newblock \bibinfo{journal}{\emph{{ACM} Trans. Math. Software}}
  \bibinfo{volume}{17} (\bibinfo{year}{1991}), \bibinfo{pages}{533--554}.
\newblock


\bibitem[Gustafsson(1994)]%
        {G1994}
\bibfield{author}{\bibinfo{person}{Kjell Gustafsson}.}
  \bibinfo{year}{1994}\natexlab{}.
\newblock \showarticletitle{Control-theoretic techniques for stepsize selection
  in implicit Runge-Kutta methods}.
\newblock \bibinfo{journal}{\emph{{ACM} Trans. Math. Software}}
  \bibinfo{volume}{20} (\bibinfo{year}{1994}), \bibinfo{pages}{496--512}.
\newblock


\bibitem[Hairer et~al\mbox{.}(2000)]%
        {HNW2000}
\bibfield{author}{\bibinfo{person}{Ernst Hairer}, \bibinfo{person}{Syvert~P.
  N{\o}rsett}, {and} \bibinfo{person}{Gerhard Wanner}.}
  \bibinfo{year}{2000}\natexlab{}.
\newblock \bibinfo{booktitle}{\emph{Solving Ordinary Differential Equations I
  -- Nonstiff Problems} (\bibinfo{edition}{2} ed.)}. \bibinfo{series}{Springer
  Series in Computational Mathematics}, Vol.~\bibinfo{volume}{8}.
\newblock \bibinfo{publisher}{Springer-Verlag}, \bibinfo{address}{Berlin}.
\newblock


\bibitem[Hairer and Wanner(2002)]%
        {HW2002}
\bibfield{author}{\bibinfo{person}{Ernst Hairer} {and} \bibinfo{person}{Gerhard
  Wanner}.} \bibinfo{year}{2002}\natexlab{}.
\newblock \bibinfo{booktitle}{\emph{Solving Ordinary Differential Equations II
  -- Stiff and Differential-Algebraic Problems} (\bibinfo{edition}{2} ed.)}.
  \bibinfo{series}{Springer Series in Computational Mathematics},
  Vol.~\bibinfo{volume}{14}.
\newblock \bibinfo{publisher}{Springer-Verlag}, \bibinfo{address}{Berlin}.
\newblock


\bibitem[Hiebert and Shampine(1980)]%
        {HS1980}
\bibfield{author}{\bibinfo{person}{Kathie~L. Hiebert} {and}
  \bibinfo{person}{Lawrence~F. Shampine}.} \bibinfo{year}{1980}\natexlab{}.
\newblock \bibinfo{booktitle}{\emph{Implicitly Defined Output Points for
  Solutions of ODEs}}.
\newblock \bibinfo{type}{{T}echnical {R}eport} SAND80-0180.
  \bibinfo{institution}{Sandia National Laboratories}.
\newblock


\bibitem[Higueras(2006)]%
        {H2006}
\bibfield{author}{\bibinfo{person}{Inmaculada Higueras}.}
  \bibinfo{year}{2006}\natexlab{}.
\newblock \showarticletitle{Strong {Stability} for {Additive} {Runge}–{Kutta}
  {Methods}}.
\newblock \bibinfo{journal}{\emph{SIAM J. Numer. Anal.}} \bibinfo{volume}{44},
  \bibinfo{number}{4} (\bibinfo{date}{Jan.} \bibinfo{year}{2006}),
  \bibinfo{pages}{1735--1758}.
\newblock
\urldef\tempurl%
\url{https://doi.org/10.1137/040612968}
\showDOI{\tempurl}


\bibitem[Higueras(2009)]%
        {H2009}
\bibfield{author}{\bibinfo{person}{Inmaculada Higueras}.}
  \bibinfo{year}{2009}\natexlab{}.
\newblock \showarticletitle{Characterizing {Strong} {Stability} {Preserving}
  {Additive} {Runge}-{Kutta} {Methods}}.
\newblock \bibinfo{journal}{\emph{J Sci Comput}} \bibinfo{volume}{39},
  \bibinfo{number}{1} (\bibinfo{date}{April} \bibinfo{year}{2009}),
  \bibinfo{pages}{115--128}.
\newblock
\urldef\tempurl%
\url{https://doi.org/10.1007/s10915-008-9252-2}
\showDOI{\tempurl}


\bibitem[Higueras et~al\mbox{.}(2014)]%
        {HHKK2014}
\bibfield{author}{\bibinfo{person}{Inmaculada Higueras},
  \bibinfo{person}{Natalie Happenhofer}, \bibinfo{person}{Othmar Koch}, {and}
  \bibinfo{person}{Friedrich Kupka}.} \bibinfo{year}{2014}\natexlab{}.
\newblock \showarticletitle{Optimized strong stability preserving {IMEX}
  {Runge}–{Kutta} methods}.
\newblock \bibinfo{journal}{\emph{J. Comput. Appl. Math.}}
  \bibinfo{volume}{272} (\bibinfo{date}{Dec.} \bibinfo{year}{2014}),
  \bibinfo{pages}{116--140}.
\newblock
\urldef\tempurl%
\url{https://doi.org/10.1016/j.cam.2014.05.011}
\showDOI{\tempurl}


\bibitem[Hindmarsh et~al\mbox{.}(2005)]%
        {HBGLSSW2005}
\bibfield{author}{\bibinfo{person}{Alan~C. Hindmarsh},
  \bibinfo{person}{Peter~N. Brown}, \bibinfo{person}{Keith~E. Grant},
  \bibinfo{person}{Steven~L. Lee}, \bibinfo{person}{Radu Serban},
  \bibinfo{person}{Dan~E. Shumaker}, {and} \bibinfo{person}{Carol~S.
  Woodward}.} \bibinfo{year}{2005}\natexlab{}.
\newblock \showarticletitle{SUNDIALS: Suite of Nonlinear and
  Differential/Algebraic Equation Solvers}.
\newblock \bibinfo{journal}{\emph{{ACM} Trans. Math. Software}}
  \bibinfo{volume}{31}, \bibinfo{number}{3} (\bibinfo{year}{2005}),
  \bibinfo{pages}{363--396}.
\newblock


\bibitem[Hindmarsh and Serban(2012)]%
        {HS2012}
\bibfield{author}{\bibinfo{person}{Alan~C. Hindmarsh} {and}
  \bibinfo{person}{Radu Serban}.} \bibinfo{year}{2012}\natexlab{}.
\newblock \bibinfo{booktitle}{\emph{User Documentation for CVODE v2.7.0}}.
\newblock \bibinfo{type}{{T}echnical {R}eport} UCRL-SM-208108.
  \bibinfo{institution}{Lawrence Livermore National Laboratory}.
\newblock


\bibitem[Kang et~al\mbox{.}(2022)]%
        {KangEtAl2022}
\bibfield{author}{\bibinfo{person}{Shinhoo Kang}, \bibinfo{person}{Alp Dener},
  \bibinfo{person}{Aidan Hamilton}, \bibinfo{person}{Hong Zhang},
  \bibinfo{person}{Emil~M. Constantinescu}, {and} \bibinfo{person}{Robert~L.
  Jacob}.} \bibinfo{year}{2022}\natexlab{}.
\newblock \bibinfo{title}{Multirate Partitioned {R}unge--{K}utta Methods for
  Coupled {N}avier--{S}tokes Equations}.
\newblock
\newblock
\urldef\tempurl%
\url{https://doi.org/10.48550/ARXIV.2202.11890}
\showDOI{\tempurl}


\bibitem[Kennedy and Carpenter(2003a)]%
        {KC2003}
\bibfield{author}{\bibinfo{person}{Christopher~A. Kennedy} {and}
  \bibinfo{person}{Mark~H. Carpenter}.} \bibinfo{year}{2003}\natexlab{a}.
\newblock \showarticletitle{Additive Runge-Kutta schemes for
  convection-diffusion-reaction equations}.
\newblock \bibinfo{journal}{\emph{Appl.~Numer.~Math.}}  \bibinfo{volume}{44}
  (\bibinfo{year}{2003}), \bibinfo{pages}{139--181}.
\newblock
\urldef\tempurl%
\url{https://doi.org/10.1016/S0168-9274(02)00138-1}
\showDOI{\tempurl}


\bibitem[Kennedy and Carpenter(2003b)]%
        {kennedyAdditiveRungeKutta2003}
\bibfield{author}{\bibinfo{person}{Christopher~A. Kennedy} {and}
  \bibinfo{person}{Mark~H. Carpenter}.} \bibinfo{year}{2003}\natexlab{b}.
\newblock \showarticletitle{Additive {{Runge}}\textendash{{Kutta}} Schemes for
  Convection\textendash Diffusion\textendash Reaction Equations}.
\newblock \bibinfo{journal}{\emph{Applied Numerical Mathematics}}
  \bibinfo{volume}{44}, \bibinfo{number}{1-2} (\bibinfo{date}{Jan.}
  \bibinfo{year}{2003}), \bibinfo{pages}{139--181}.
\newblock
\showISSN{01689274}
\urldef\tempurl%
\url{https://doi.org/10.1016/S0168-9274(02)00138-1}
\showDOI{\tempurl}


\bibitem[Kennedy and Carpenter(2019a)]%
        {KC2019}
\bibfield{author}{\bibinfo{person}{Christopher~A. Kennedy} {and}
  \bibinfo{person}{Mark~H. Carpenter}.} \bibinfo{year}{2019}\natexlab{a}.
\newblock \showarticletitle{Higher-order additive Runge-Kutta schemes for
  ordinary differential equations}.
\newblock \bibinfo{journal}{\emph{Appl.~Numer.~Math.}}  \bibinfo{volume}{136}
  (\bibinfo{year}{2019}), \bibinfo{pages}{183--205}.
\newblock


\bibitem[Kennedy and Carpenter(2019b)]%
        {kennedyHigherorderAdditiveRunge2019}
\bibfield{author}{\bibinfo{person}{Christopher~A. Kennedy} {and}
  \bibinfo{person}{Mark~H. Carpenter}.} \bibinfo{year}{2019}\natexlab{b}.
\newblock \showarticletitle{Higher-Order Additive {{Runge}}\textendash{{Kutta}}
  Schemes for Ordinary Differential Equations}.
\newblock \bibinfo{journal}{\emph{Applied Numerical Mathematics}}
  \bibinfo{volume}{136} (\bibinfo{date}{Feb.} \bibinfo{year}{2019}),
  \bibinfo{pages}{183--205}.
\newblock
\showISSN{01689274}
\urldef\tempurl%
\url{https://doi.org/10.1016/j.apnum.2018.10.007}
\showDOI{\tempurl}


\bibitem[Knoth and Wolke(1998)]%
        {KW1998}
\bibfield{author}{\bibinfo{person}{Oswald Knoth} {and} \bibinfo{person}{Ralf
  Wolke}.} \bibinfo{year}{1998}\natexlab{}.
\newblock \showarticletitle{Implicit-explicit Runge-Kutta methods for computing
  atmospheric reactive flows}.
\newblock \bibinfo{journal}{\emph{Appl. Numer. Math.}} \bibinfo{volume}{28},
  \bibinfo{number}{2} (\bibinfo{year}{1998}), \bibinfo{pages}{327--341}.
\newblock


\bibitem[Kreckel et~al\mbox{.}(2010)]%
        {Kreckel2010}
\bibfield{author}{\bibinfo{person}{H. Kreckel}, \bibinfo{person}{H. Bruhns},
  \bibinfo{person}{M. {\v C}{\'\i}{\v z}ek}, \bibinfo{person}{S.~C.~O. Glover},
  \bibinfo{person}{K.~A. Miller}, \bibinfo{person}{X. Urbain}, {and}
  \bibinfo{person}{D.~W. Savin}.} \bibinfo{year}{2010}\natexlab{}.
\newblock \showarticletitle{Experimental Results for H2 Formation from H- and H
  and Implications for First Star Formation}.
\newblock \bibinfo{journal}{\emph{Science}} \bibinfo{volume}{329},
  \bibinfo{number}{5987} (\bibinfo{year}{2010}), \bibinfo{pages}{69--71}.
\newblock
\showISSN{0036-8075}
\urldef\tempurl%
\url{https://doi.org/10.1126/science.1187191}
\showDOI{\tempurl}


\bibitem[Luan et~al\mbox{.}(2020)]%
        {luanNewClassHighOrder2020}
\bibfield{author}{\bibinfo{person}{Vu~Thai Luan}, \bibinfo{person}{Rujeko
  Chinomona}, {and} \bibinfo{person}{Daniel~R. Reynolds}.}
  \bibinfo{year}{2020}\natexlab{}.
\newblock \showarticletitle{A {{New Class}} of {{High}}-{{Order Methods}} for
  {{Multirate Differential Equations}}}.
\newblock \bibinfo{journal}{\emph{SIAM Journal on Scientific Computing}}
  \bibinfo{volume}{42}, \bibinfo{number}{2} (\bibinfo{date}{Jan.}
  \bibinfo{year}{2020}), \bibinfo{pages}{A1245--A1268}.
\newblock
\showISSN{1064-8275, 1095-7197}
\urldef\tempurl%
\url{https://doi.org/10.1137/19M125621X}
\showDOI{\tempurl}


\bibitem[Ober(2022)]%
        {tempus-website}
\bibfield{author}{\bibinfo{person}{Curtis~C. Ober}.}
  \bibinfo{year}{2022}\natexlab{}.
\newblock \bibinfo{title}{The {T}empus {P}roject {W}ebsite}.
\newblock \bibinfo{howpublished}{\url{https://trilinos.github.io/tempus.html}}.
\newblock


\bibitem[Pareschi and Russo(2005)]%
        {PR2005}
\bibfield{author}{\bibinfo{person}{Lorenzo Pareschi} {and}
  \bibinfo{person}{Giovanni Russo}.} \bibinfo{year}{2005}\natexlab{}.
\newblock \showarticletitle{Implicit–{Explicit} {Runge}–{Kutta} {Schemes}
  and {Applications} to {Hyperbolic} {Systems} with {Relaxation}}.
\newblock \bibinfo{journal}{\emph{J Sci Comput}} \bibinfo{volume}{25},
  \bibinfo{number}{1} (\bibinfo{date}{Oct.} \bibinfo{year}{2005}),
  \bibinfo{pages}{129--155}.
\newblock
\urldef\tempurl%
\url{https://doi.org/10.1007/s10915-004-4636-4}
\showDOI{\tempurl}


\bibitem[Rackauckas and Nie(2017)]%
        {rackauckas2017differentialequations}
\bibfield{author}{\bibinfo{person}{Christopher Rackauckas} {and}
  \bibinfo{person}{Qing Nie}.} \bibinfo{year}{2017}\natexlab{}.
\newblock \showarticletitle{{DifferentialEquations.jl--a performant and
  feature-rich ecosystem for solving differential equations in Julia}}.
\newblock \bibinfo{journal}{\emph{Journal of Open Research Software}}
  \bibinfo{volume}{5}, \bibinfo{number}{1} (\bibinfo{year}{2017}),
  \bibinfo{pages}{15}.
\newblock


\bibitem[Radhakrishnan et~al\mbox{.}(2016)]%
        {radhakrishnan2016phase}
\bibfield{author}{\bibinfo{person}{Bala Radhakrishnan}, \bibinfo{person}{Sarma
  Gorti}, {and} \bibinfo{person}{Suresh~Sudharsanam Babu}.}
  \bibinfo{year}{2016}\natexlab{}.
\newblock \showarticletitle{Phase field simulations of autocatalytic formation
  of alpha lamellar colonies in Ti-6Al-4V}.
\newblock \bibinfo{journal}{\emph{Metallurgical and Materials Transactions A}}
  \bibinfo{volume}{47}, \bibinfo{number}{12} (\bibinfo{year}{2016}),
  \bibinfo{pages}{6577--6592}.
\newblock


\bibitem[Ranocha et~al\mbox{.}(2021)]%
        {ranocha2021optimized}
\bibfield{author}{\bibinfo{person}{Hendrik Ranocha}, \bibinfo{person}{Lisandro
  Dalcin}, \bibinfo{person}{Matteo Parsani}, {and} \bibinfo{person}{David~I.
  Ketcheson}.} \bibinfo{year}{2021}\natexlab{}.
\newblock \showarticletitle{Optimized Runge-Kutta Methods with Automatic Step
  Size Control for Compressible Computational Fluid Dynamics}.
\newblock \bibinfo{journal}{\emph{Communications on Applied Mathematics and
  Computation}} (\bibinfo{year}{2021}), \bibinfo{pages}{1--38}.
\newblock


\bibitem[Reynolds et~al\mbox{.}(2022a)]%
        {RGB2021}
\bibfield{author}{\bibinfo{person}{Daniel~R. Reynolds},
  \bibinfo{person}{David~J. Gardner}, {and} \bibinfo{person}{Cody~J. Balos}.}
  \bibinfo{year}{2022}\natexlab{a}.
\newblock \bibinfo{title}{{SUNDIALS} {ManyVector}+Multirate Demonstration
  Code}.
\newblock
  \bibinfo{howpublished}{\url{https://github.com/sundials-codes/sundials-manyvector-demo}}.
\newblock


\bibitem[Reynolds et~al\mbox{.}(2019)]%
        {reynoldsSUNDIALSMultiphysicsMPIManyVector2019a}
\bibfield{author}{\bibinfo{person}{Daniel~R. Reynolds},
  \bibinfo{person}{David~J. Gardner}, \bibinfo{person}{Cody~J. Balos}, {and}
  \bibinfo{person}{Carol~S. Woodward}.} \bibinfo{year}{2019}\natexlab{}.
\newblock \showarticletitle{{{SUNDIALS Multiphysics}}+{{MPIManyVector
  Performance Testing}}}.
\newblock \bibinfo{journal}{\emph{arXiv:1909.12966 [cs]}}
  (\bibinfo{date}{Sept.} \bibinfo{year}{2019}).
\newblock
\showeprint[arxiv]{1909.12966}~[cs]


\bibitem[Reynolds et~al\mbox{.}(2022b)]%
        {arkodeUserGuide}
\bibfield{author}{\bibinfo{person}{Daniel~R. Reynolds},
  \bibinfo{person}{David~J. Gardner}, \bibinfo{person}{Carol~S. Woodward},
  {and} \bibinfo{person}{Cody~J. Balos}.} \bibinfo{year}{2022}\natexlab{b}.
\newblock \bibinfo{title}{User Documentation for ARKODE}.
\newblock
  \bibinfo{howpublished}{\url{https://sundials.readthedocs.io/en/latest/arkode}}.
\newblock


\bibitem[Roberts et~al\mbox{.}(2021)]%
        {roberts2021implicit}
\bibfield{author}{\bibinfo{person}{Steven Roberts}, \bibinfo{person}{John
  Loffeld}, \bibinfo{person}{Arash Sarshar}, \bibinfo{person}{Carol~S
  Woodward}, {and} \bibinfo{person}{Adrian Sandu}.}
  \bibinfo{year}{2021}\natexlab{}.
\newblock \showarticletitle{Implicit multirate {GARK} methods}.
\newblock \bibinfo{journal}{\emph{Journal of Scientific Computing}}
  \bibinfo{volume}{87}, \bibinfo{number}{1} (\bibinfo{year}{2021}),
  \bibinfo{pages}{1--32}.
\newblock


\bibitem[Roberts et~al\mbox{.}(2020)]%
        {robertsCoupledMultirateInfinitesimal2020}
\bibfield{author}{\bibinfo{person}{Steven Roberts}, \bibinfo{person}{Arash
  Sarshar}, {and} \bibinfo{person}{Adrian Sandu}.}
  \bibinfo{year}{2020}\natexlab{}.
\newblock \showarticletitle{Coupled Multirate Infinitesimal GARK Schemes for
  Stiff Systems with Multiple Time Scales}.
\newblock \bibinfo{journal}{\emph{SIAM Journal on Scientific Computing}}
  \bibinfo{volume}{42}, \bibinfo{number}{3} (\bibinfo{year}{2020}),
  \bibinfo{pages}{A1609--A1638}.
\newblock
\urldef\tempurl%
\url{https://doi.org/10.1137/19M1266952}
\showDOI{\tempurl}


\bibitem[Sandu(2019)]%
        {sanduClassMultirateInfinitesimal2019}
\bibfield{author}{\bibinfo{person}{Adrian Sandu}.}
  \bibinfo{year}{2019}\natexlab{}.
\newblock \showarticletitle{A {{Class}} of {{Multirate Infinitesimal GARK
  Methods}}}.
\newblock \bibinfo{journal}{\emph{SIAM J. Numer. Anal.}} \bibinfo{volume}{57},
  \bibinfo{number}{5} (\bibinfo{date}{Jan.} \bibinfo{year}{2019}),
  \bibinfo{pages}{2300--2327}.
\newblock
\showISSN{0036-1429, 1095-7170}
\urldef\tempurl%
\url{https://doi.org/10.1137/18M1205492}
\showDOI{\tempurl}


\bibitem[Schlegel et~al\mbox{.}(2009a)]%
        {schlegelMultirateRungeKutta2009}
\bibfield{author}{\bibinfo{person}{Martin Schlegel}, \bibinfo{person}{Oswald
  Knoth}, \bibinfo{person}{Martin Arnold}, {and} \bibinfo{person}{Ralf Wolke}.}
  \bibinfo{year}{2009}\natexlab{a}.
\newblock \showarticletitle{Multirate {{Runge}}\textendash{{Kutta}} Schemes for
  Advection Equations}.
\newblock \bibinfo{journal}{\emph{J. Comput. Appl. Math.}}
  \bibinfo{volume}{226}, \bibinfo{number}{2} (\bibinfo{date}{April}
  \bibinfo{year}{2009}), \bibinfo{pages}{345--357}.
\newblock
\showISSN{0377-0427}
\urldef\tempurl%
\url{https://doi.org/10.1016/j.cam.2008.08.009}
\showDOI{\tempurl}


\bibitem[Schlegel et~al\mbox{.}(2009b)]%
        {SKAW2009}
\bibfield{author}{\bibinfo{person}{Martin Schlegel}, \bibinfo{person}{Oswald
  Knoth}, \bibinfo{person}{Martin Arnold}, {and} \bibinfo{person}{Ralf Wolke}.}
  \bibinfo{year}{2009}\natexlab{b}.
\newblock \showarticletitle{Multirate Runge–Kutta schemes for advection
  equations}.
\newblock \bibinfo{journal}{\emph{J. Comput. Appl. Math.}}
  \bibinfo{volume}{226} (\bibinfo{year}{2009}), \bibinfo{pages}{345--357}.
\newblock


\bibitem[Schlegel et~al\mbox{.}(2012a)]%
        {SKAW2012a}
\bibfield{author}{\bibinfo{person}{Martin Schlegel}, \bibinfo{person}{Oswald
  Knoth}, \bibinfo{person}{Martin Arnold}, {and} \bibinfo{person}{Ralf Wolke}.}
  \bibinfo{year}{2012}\natexlab{a}.
\newblock \showarticletitle{Implementation of multirate time integration
  methods for air pollution modelling}.
\newblock \bibinfo{journal}{\emph{GMD}}  \bibinfo{volume}{5}
  (\bibinfo{year}{2012}), \bibinfo{pages}{1395--1405}.
\newblock


\bibitem[Schlegel et~al\mbox{.}(2012b)]%
        {SKAW2012b}
\bibfield{author}{\bibinfo{person}{Martin Schlegel}, \bibinfo{person}{Oswald
  Knoth}, \bibinfo{person}{Martin Arnold}, {and} \bibinfo{person}{Ralf Wolke}.}
  \bibinfo{year}{2012}\natexlab{b}.
\newblock \showarticletitle{Numerical solution of multiscale problems in
  atmospheric modeling}.
\newblock \bibinfo{journal}{\emph{Appl. Numer. Math.}}  \bibinfo{volume}{62}
  (\bibinfo{year}{2012}), \bibinfo{pages}{1531--1542}.
\newblock


\bibitem[Shu(2003)]%
        {Shu2003}
\bibfield{author}{\bibinfo{person}{Chi-Wang Shu}.}
  \bibinfo{year}{2003}\natexlab{}.
\newblock \showarticletitle{High-order Finite Difference and Finite Volume WENO
  Schemes and Discontinuous Galerkin Methods for CFD}.
\newblock \bibinfo{journal}{\emph{International Journal of Computational Fluid
  Dynamics}} \bibinfo{volume}{17}, \bibinfo{number}{2} (\bibinfo{year}{2003}),
  \bibinfo{pages}{107--118}.
\newblock
\urldef\tempurl%
\url{https://doi.org/10.1080/1061856031000104851}
\showDOI{\tempurl}


\bibitem[Sitaraman et~al\mbox{.}(2021)]%
        {sitaraman2021adaptive}
\bibfield{author}{\bibinfo{person}{Hariswaran Sitaraman},
  \bibinfo{person}{Shashank Yellapantula}, \bibinfo{person}{Marc T.~Henry de
  Frahan}, \bibinfo{person}{Bruce Perry}, \bibinfo{person}{Jon Rood},
  \bibinfo{person}{Ray Grout}, {and} \bibinfo{person}{Marc Day}.}
  \bibinfo{year}{2021}\natexlab{}.
\newblock \showarticletitle{Adaptive mesh based combustion simulations of
  direct fuel injection effects in a supersonic cavity flame-holder}.
\newblock \bibinfo{journal}{\emph{Combustion and Flame}}  \bibinfo{volume}{232}
  (\bibinfo{year}{2021}), \bibinfo{pages}{111531}.
\newblock


\bibitem[S{\"o}derlind(1998)]%
        {S1998}
\bibfield{author}{\bibinfo{person}{Gustaf S{\"o}derlind}.}
  \bibinfo{year}{1998}\natexlab{}.
\newblock \showarticletitle{The automatic control of numerical integration}.
\newblock \bibinfo{journal}{\emph{CWI Quarterly}}  \bibinfo{volume}{11}
  (\bibinfo{year}{1998}), \bibinfo{pages}{55--74}.
\newblock


\bibitem[S{\"o}derlind(2003)]%
        {S2003}
\bibfield{author}{\bibinfo{person}{Gustaf S{\"o}derlind}.}
  \bibinfo{year}{2003}\natexlab{}.
\newblock \showarticletitle{Digital filters in adaptive time-stepping}.
\newblock \bibinfo{journal}{\emph{{ACM} Trans. Math. Software}}
  \bibinfo{volume}{29} (\bibinfo{year}{2003}), \bibinfo{pages}{1--26}.
\newblock


\bibitem[S{\"o}derlind(2006)]%
        {S2006}
\bibfield{author}{\bibinfo{person}{Gustaf S{\"o}derlind}.}
  \bibinfo{year}{2006}\natexlab{}.
\newblock \showarticletitle{Time-step selection algorithms: Adaptivity, control
  and signal processing}.
\newblock \bibinfo{journal}{\emph{Appl.~Numer.~Math.}}  \bibinfo{volume}{56}
  (\bibinfo{year}{2006}), \bibinfo{pages}{488--502}.
\newblock


\bibitem[Tomov et~al\mbox{.}(2010)]%
        {tdb10}
\bibfield{author}{\bibinfo{person}{Stanimire Tomov}, \bibinfo{person}{Jack
  Dongarra}, {and} \bibinfo{person}{Marc Baboulin}.}
  \bibinfo{year}{2010}\natexlab{}.
\newblock \showarticletitle{{Towards dense linear algebra for hybrid GPU
  accelerated manycore systems}}.
\newblock \bibinfo{journal}{\emph{Parallel Comput.}} \bibinfo{volume}{36},
  \bibinfo{number}{5-6} (\bibinfo{date}{June} \bibinfo{year}{2010}),
  \bibinfo{pages}{232--240}.
\newblock
\showISSN{0167-8191}
\urldef\tempurl%
\url{https://doi.org/10.1016/j.parco.2009.12.005}
\showDOI{\tempurl}


\bibitem[Trevisan and Tennyson(2002)]%
        {Trevisan_2002}
\bibfield{author}{\bibinfo{person}{Cynthia~S. Trevisan} {and}
  \bibinfo{person}{Jonathan Tennyson}.} \bibinfo{year}{2002}\natexlab{}.
\newblock \showarticletitle{Calculated rates for the electron impact
  dissociation of molecular hydrogen, deuterium and tritium}.
\newblock \bibinfo{journal}{\emph{Plasma Physics and Controlled Fusion}}
  \bibinfo{volume}{44}, \bibinfo{number}{7} (\bibinfo{date}{Jun}
  \bibinfo{year}{2002}), \bibinfo{pages}{1263--1276}.
\newblock
\urldef\tempurl%
\url{https://doi.org/10.1088/0741-3335/44/7/315}
\showDOI{\tempurl}


\bibitem[unit~project group(2013)]%
        {IVPTestSet}
\bibfield{author}{\bibinfo{person}{INdAM~Bari unit~project group}.}
  \bibinfo{year}{2013}\natexlab{}.
\newblock \bibinfo{title}{Test Set for {IVP} Solvers}.
\newblock
  \bibinfo{howpublished}{\url{https://archimede.dm.uniba.it/~testset/testsetivpsolvers}}.
\newblock


\bibitem[Vogl et~al\mbox{.}(2019)]%
        {vogl2019evaluation}
\bibfield{author}{\bibinfo{person}{Christopher~J. Vogl},
  \bibinfo{person}{Andrew Steyer}, \bibinfo{person}{Daniel~R. Reynolds},
  \bibinfo{person}{Paul~A. Ullrich}, {and} \bibinfo{person}{Carol~S.
  Woodward}.} \bibinfo{year}{2019}\natexlab{}.
\newblock \showarticletitle{Evaluation of Implicit-Explicit Additive
  Runge-Kutta Integrators for the HOMME-NH Dynamical Core}.
\newblock \bibinfo{journal}{\emph{J. Adv. Model. Earth Syst.}}
  \bibinfo{volume}{11}, \bibinfo{number}{12} (\bibinfo{year}{2019}),
  \bibinfo{pages}{4228--4244}.
\newblock


\bibitem[Wensch et~al\mbox{.}(2009)]%
        {wenschMultirateInfinitesimalStep2009}
\bibfield{author}{\bibinfo{person}{J{\"o}rg Wensch}, \bibinfo{person}{Oswald
  Knoth}, {and} \bibinfo{person}{Alexander Galant}.}
  \bibinfo{year}{2009}\natexlab{}.
\newblock \showarticletitle{Multirate Infinitesimal Step Methods for
  Atmospheric Flow Simulation}.
\newblock \bibinfo{journal}{\emph{Bit Numer Math}} \bibinfo{volume}{49},
  \bibinfo{number}{2} (\bibinfo{date}{June} \bibinfo{year}{2009}),
  \bibinfo{pages}{449--473}.
\newblock
\showISSN{0006-3835, 1572-9125}
\urldef\tempurl%
\url{https://doi.org/10.1007/s10543-009-0222-3}
\showDOI{\tempurl}


\bibitem[Zonneveld(1963)]%
        {Z1963}
\bibfield{author}{\bibinfo{person}{J.A. Zonneveld}.}
  \bibinfo{year}{1963}\natexlab{}.
\newblock \bibinfo{booktitle}{\emph{Automatic integration of ordinary
  differential equations}}.
\newblock \bibinfo{type}{{T}echnical {R}eport} R743.
  \bibinfo{institution}{Mathematisch Centrum}, \bibinfo{address}{Postbus 4079,
  1009AB Amsterdam}.
\newblock


\end{thebibliography}




\end{document}